\documentclass[aps,pre,twocolumn]{revtex4} 
\usepackage{hyperref,natbib,doi}
\usepackage{graphicx,caption}
\usepackage{amsmath}
\usepackage{dcolumn}
\usepackage[utf8]{inputenc}
\usepackage{amssymb}
\begin{document}
\title{The effect of particle gas composition and 
boundary conditions on triboplasma generation: a computational study using the particle-in-cell method}
\author{D. Tsiklauri$^1$, S. Karabasov$^2$, V. Prodaevich$^3$}
\affiliation{$^1$ School of Physics and Astronomy, Queen Mary University of London, London, E1 4NS, United Kingdom \\
$^2$ School of Engineering and Material Sciences, Queen Mary University of London, London, E1 4NS, United Kingdom \\
$^3$ Engineering Company Eco-Ardens, 47 Gogol street, Nizhny Novgorod, 603109, Russia }
\begin{abstract}
Two dimensional particle in cell simulations of free 
charge creation by collisional ionization of C12 and C60 
molecules immersed in plasma for the parameters of 
relevance to plasma gasification are presented.
Our main findings are that
(i) in uniform plasmas with smooth walls two optimal values 
which emerge for free electron production by collisional ionization 
(i.e. a most efficient discharge condition creation) are $C60:C12$ fractions of 
$10:90$ and $80:20$,
(ii) in plasmas with rough walls, modelled by comb-like electric field
at the boundary, the case of tangential electric field 
creates significant charge localization in C12+ and C60+ species, 
again creating most favorable discharge condition for 
tribo-electrically generated plasma.
The numerical simulation results are discussed with reference 
to recent triboelectric plasma experiments and are 
corroborated by suitable analytical models.
\end{abstract}

\maketitle

\section{Introduction}

With the depletion of fossil resources and the mounting quantity of 
waste, energy generation and waste disposal have become very important 
problems of modern society. Waste-to-energy (WTE) approaches which aim 
to generate energy as heat or power from waste can provide a balanced 
solution to both these problems. One of the most promising WTE technologies 
is associated with the recuperation of energy via transforming 
non-recyclable materials through a combination of different high 
temperature-involving procedures such as waste gasification and pyrolysis. 
The advantages of the above thermal techniques over the conventional 
WTE techniques such as incineration and combustion include higher 
recycling rates, lower toxic gas emissions, higher energy efficiencies, 
lower costs, smaller carbon footprints and reduced environmental 
impact \cite{ref1}. Importantly, gasification converts solid waste 
into a highly fungible synthetic gas (or syngas) very rich in hydrogen 
and carbon monoxide, that can be converted into clean electricity or 
other high value fuels/chemicals, including methanol, SNG (synthetic natural gas) 
or pure hydrogen  \cite{ref2,ref3,ref4}. 

The use of plasma power has been popular within thermal waste treatments 
for its ability to completely decompose the input waste material into a 
tar-free synthetic gas and an inert, environmentally stable, vitreous 
material (slag) and preparing the syngas for efficient electricity 
production or catalytic transformation \cite{ref6}. Because of the 
potential advantages, plasma technologies have been developed for the 
destruction and removal of various hazardous wastes, such as polychlorinated 
biphenyls  (PCBs) \cite{ref7}, medical waste \cite{ref8}, metallurgical wastes, 
incineration fly ash \cite{ref9}, and low-level radioactive wastes.

In addition to the waste gasification, plasma assisted combustion is a 
very active topic of research on its own right, which covers the topic 
ignition enhancement, ultra-lean combustion, cool flames, flameless combustion, 
and controllability of plasma discharge \cite{ref10}.

Currently, in many engineering applications plasma have been generated
 by constant current or electromagnetic field for which an external supply 
 of electric energy is needed. For example, in the existing plasma gasification 
 technologies, only additions of combustion heat supplied by the waste 
 feedstock or a fuel additive make the process suited to large waste streams \cite{ref11}. 
The main cost of the current plasma power technologies is associated with 
the energy required to artificially create significant electromagnetic or 
electrostatic fields to trigger and sustain gas discharges. 
For example, for dry air, few MVm$^{-1}$ is required to trigger a corona 
discharge \cite{ref12}. This is of the same order of magnitude as the maximum 
power generated by the conventional high-energy particle accelerators. Such 
limit is due to the radio frequency (RF) breakdown phenomenon: when such large 
electric field is used in the accelerator cavities, it causes accelerator to 
be effectively short-circuited in accordance with the so-called Kilpatrick 
limit \cite{k57}. Notably, this limit is overcome in novel accelerators 
which are based on plasmas, the so-called {\it plasma wake field acceleration}, 
and which can sustain electric fields up to tens of GVm$^{-1}$, without 
electric short-circuiting \cite{me1,me2}. 

{ To trigger the discharge, hence creating plasma,
electrodes powered by direct current are typically used \cite{ref13} }. 
Altogether, the high cost of conventional plasma generators and short 
working life-time of electrodes (circa 500 hours) encourages researchers to 
consider other sources of plasma generation which do not need either the 
external electro-magnetic field or the electrodes.

Triboelectricity, or electricity generation by {\it mechanical friction}, 
can provide such an alternative source of plasma generation. 
An example of triboelectric charging in nature is the ash produced from the 
volcanic eruption that collides with one another producing significant 
charging which is discharged through lightning strikes. 

Triboelectric plasma generation to ultimately replace the 
expensive direct current operated plasma torches can greatly 
improve efficiency of modern waste-to-energy gasification schemes 
while maintaining a very low emission signature. 

In a recent laboratory experiment \cite{kar1} the generation of 
plasma through a triboelectric effect was reported by impinging a 
high-speed (150-200 m/s) microjet of deionised water on a dialectric surface. 
A naturally formed, stable, unconstrained and topologically coherent 
triboplasma region in the form of a coherent toroidal structure was 
obtained in atmospheric pressure conditions without any external 
electromagnetic action.

\begin{figure}
\centering
\includegraphics[width=0.5\textwidth]{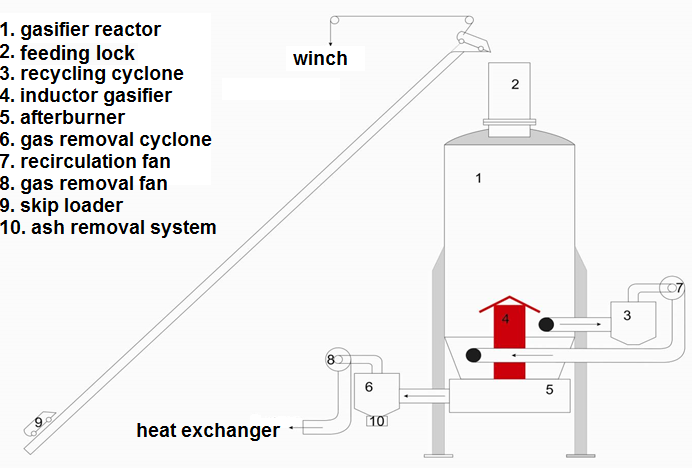}
\caption{Schematic of the triboplasma gasifier apparatus of 
Engineering Company Eco-Ardens.}
\label{f1}
\end{figure}

\begin{figure}
\centering
\includegraphics[width=0.5\textwidth]{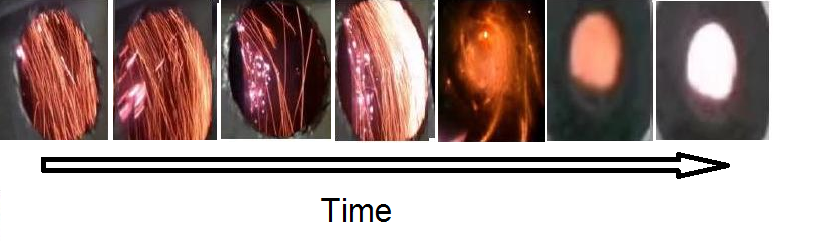}
\caption{A rendered series of snapshots through the viewing window of the 
tribo-plasma gasifier apparatus of Engineering Company Eco-Ardens. 
From left to right:
  an initial to a developed stage of the tribo-plasma generation. 
 The window location is in the centre of the reactor above the 'inductor'.}
\label{f2}
\end{figure}

An example of triboelectric plasma generation technology is the 
gasifier apparatus pioneered by LCC Engineering. In this case, a 
triboplasma region is generated by collision of ash particles 
(mostly carbon-based) in a swirling hydrodynamic flow generated by 
two tangential flow streams at a moderate flow speed (50 m/s) that 
grazes a serrated surface of an insulated steel wall. A schematic of 
the LCC Engineering apparatus is shown in Fig.\ref{f1}. 
Here the organic fuel (e.g. chicken farm waste) is supplied from 
the top of the reactor and dropped through the triboplasma zone in the 
centre (above the 'inductor gasifier' on the schematic). Because of a very 
high temperature in the centre of the reactor chamber, the waste is 
very efficiently decomposed into useful syngas, ash and a chemically 
inert slag, with virtually zero emission of toxic gases. 
Under the effect of particle-wall and particle-particle collisions, 
spark discharges are triggered, whose intensity grows as the tribo-electric 
self-charging of gas particles increases until a self-sustained localised 
tribo-plasma region emerges in the reaction zone (Fig.\ref{f2}). 

Among several fundamental issues, which remain to be 
understood before the tribolectric plasma generation can be 
used in real-life gasification applications, the following two questions stand out:

\begin{itemize} 
\item 
According to the preliminary experimental results of 
Engineering Company Eco-Ardens, gas 
samples which correspond to successful triboelectric plasma generation 
are also rich in fullerenes (C60). Fullerenes are nano-size, football-like, 
carbon molecules which have a low ionization potential in comparison with 
hydrocarbons and a large surface-to-volume ratio in comparison with 
macroscopic soot particles. They are known to be readily generated 
in carbon plasmas through a non-equilibrium growth process that involves 
dehydrogenation of hydrocarbons, nucleation of large carbon cages and 
carbon cage evaporation to produce the small highly symmetry fullerenes 
such as C60 or C70. 
The particular question addressed in this work - what is effect of 
fullerenes on triboelectric charging of a gas mixture that includes 
both C60 and the standard carbon (C12) molecules?

\item The triboelectric charge generation is dependent not only on the 
particulates present but also on the gas particle collisions with 
the uneven surfaces of the insulated conducting walls. How do the 
particle interactions with the wall can lead to an intensification 
of the triboelectric plasma effect? 
\end{itemize}

To address the above questions, we use the state-of-the art EPOCH, 
fully kinetic particle-in-cell (PIC) code for solving kinetic plasma equations 
with a self-consistent field formulation \cite{ref16}. To keep the particle 
in plasma simulations computationally feasible, a two-dimensional model of 
initially neutral carbon particles, C12 and C60, which are immersed into a 
fully ionized hydrogen plasma, is considered. In this model, the 
computational domain is covered by a Eulerian computational grid 
where the electromagnetic field equations are solved with capturing 
the characteristic Debye length. Clusters of neutral and charged carbon 
species as well as the free electrons are represented by Lagrangian 
particles which collide with including all relevant collision effects. 
The collision results in a new charge generation which is analysed for 
different concentrations of C60 and different boundary conditions to 
simulate the effect in the LCC Engineering experiment.

By analyzing the collisional ionization process it is shown that the 
rate of the new charge production from collisional ionization between 
carbon particles becomes greatly amplified once the concentration of 
fullerenes added to the gas exceeds a certain threshold value (circa 
10\% by number per volume fraction). Additional simulations reveal that 
the introduction of a non-periodic boundary condition imitating a serrated 
conducting wall of the experiment leads to a non-uniform concentration 
of the carbon particles and enhances the collision process thereby 
further enabling the increase of electric charge generation in the volume. 

\section{The model and results}

\subsection{Methods}

Two key phenomena which affect the triboelectric plasma generation 
are particle collisions and ionization. Hence, below we briefly discuss 
how these effects are implemented in the Particle-In-Cell EPOCH model \cite{ref16}.
It can be noted that many PIC models neglect particle collisions over very short 
(less than grid scale) ranges. At temperatures ($\gtrsim $ 1 keV) and 
number densities  ($\lesssim  10^{27} $m$^{-3}$) collisional effects in 
plasmas are generally considered negligible. This implies that the mean 
time between collisions is comparable to the time scales of interest, 
and the collisionless approximation used in PIC codes is valid. However, at 
lower temperatures and/or higher densities the effect of sub-grid scale 
interactions on the evolution of the system can become non-negligible.

The maximum temperature in plasma gasifiers can reach a few $10^4$ K (recall 
that 1 keV corresponds to $1.16\times10^7$ K), i.e. of the order if 1 eV. 
Typical plasma gasifier density is not readily available, but according to 
NRL Plasma Formulary, high pressure arcs have number densities of $10^{22-24}$ m$^{-3}$. 
Hence, collisional effects for gasifier setting are important.

A binary collision algorithm, based on the approach of Sentoku and Kemp has been 
implemented in EPOCH. To simplify momentum conservation treatment, collisions are 
calculated in the centre-of-momentum reference  frame of the two particles. 
Lorentz transformations are included in order to evaluate the particles' momenta 
in the centre-of-momentum frame. This ensures that in EPOCH collision algorithm is 
fully relativistic. EPOCH includes a number of different ionization models. 
These account for the different modes by which electrons ionize in both the 
external field (e.g. of an intense laser) and through collisions. 
To switch on collisions and collisional ionization in EPOCH an input file is used 
(input.deck). Four species included in the simulation are electrons, protons, C12, 
and C60. C12 has two possible ionization energies, 11.26 eV and 24.38 eV. The 
particles are immersed into a fully ionized plasma at temperature T=$10^5$ K where 
the number density of electrons and protons is set to $n=10^{15}$ m$^{-3}$. 

End simulation time in most runs where there is no electric field forcing applied at 
the boundary is set to $t_{end}=11000 / \omega_{pe}$ (Fig. \ref{f3}-\ref{f7}). 
In the case of driving electric field at the boundary (Fig. \ref{f9}-\ref{f11}) 
to simulate the wall effect, 
the end time is longer, $t_{end}=20000 / \omega_{pe}$ to make sure that the solution 
reaches a more-or-less statistically converged state at least for some of the 
considered forcing regimes.

First, the simulations are performed in a homogenous domain, without accounting 
for the effect of the uneven wall of the gasifier. The boundary conditions are 
periodic in the x-direction for
both the Electro-Magnetic (EM) fields and the particles and (ii) conducting in 
the y-direction for the fields and reflecting for the particles.
Different grid cell and particle density resolutions, as well as the domain sizes, 
are considered (Fig. \ref{f3}). Simulation results shown in figs. \ref{f4}-\ref{f8} 
correspond to the grid resolution of $288 \times 72$ with each cell being 2 Debye length, 
i.e. $\Delta=2 \lambda_D$, with $\lambda_D=V_{th,e}/\omega_{pe}$.
The concentration of C12 and C60 species is varied so that total density 
stays the same. For example, $C60:C12$ fraction of $99:1$ means that $n_{C60}=n \times 99/100$ 
while $n_{C12}=n \times 1/100$, with $n$ being $n=10^{15}$ m$^{-3}$ number density for both 
electrons and protons. Similarly, $C60:C12$ fraction of $25:75$ means that $n_{C60}=n \times 25/100$ 
while $n_{C12}=n \times 75/100$, and so on.

Secondly, the case of driving the electric field on the bottom 
wall boundary is considered. 
Periodic boundary conditions in the x- and y- directions for both the EM fields 
and the particles are imposed. The latter choice is to make sure that driving of 
the periodic boundary condition to a prescribed forcing field is fully consistent 
with the governing discretisation of Maxwell's equations. Results of these 
simulations are shown in figs. \ref{f9}-\ref{f11} which have grid 
resolution of $144 \times 36$, with each cell being 
$\Delta=4 \lambda_D$. The increased cell size is in order to have the same 
size of the computational domain in most simulation cases, with or 
without the electric field forcing.

\subsection{Homogeneous particle interaction problem}

\begin{figure}
\centering
\includegraphics[width=0.5\textwidth]{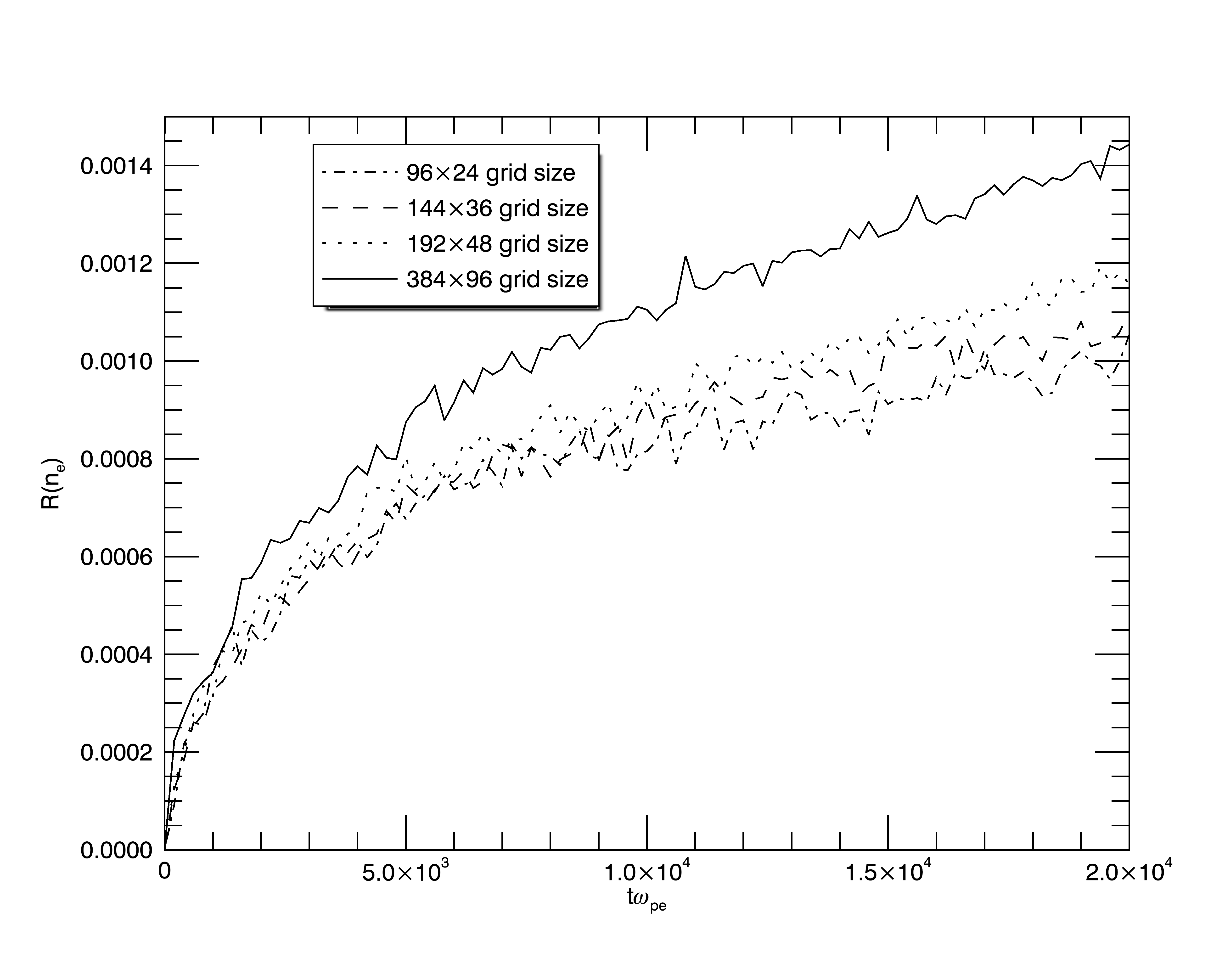}
\includegraphics[width=0.5\textwidth]{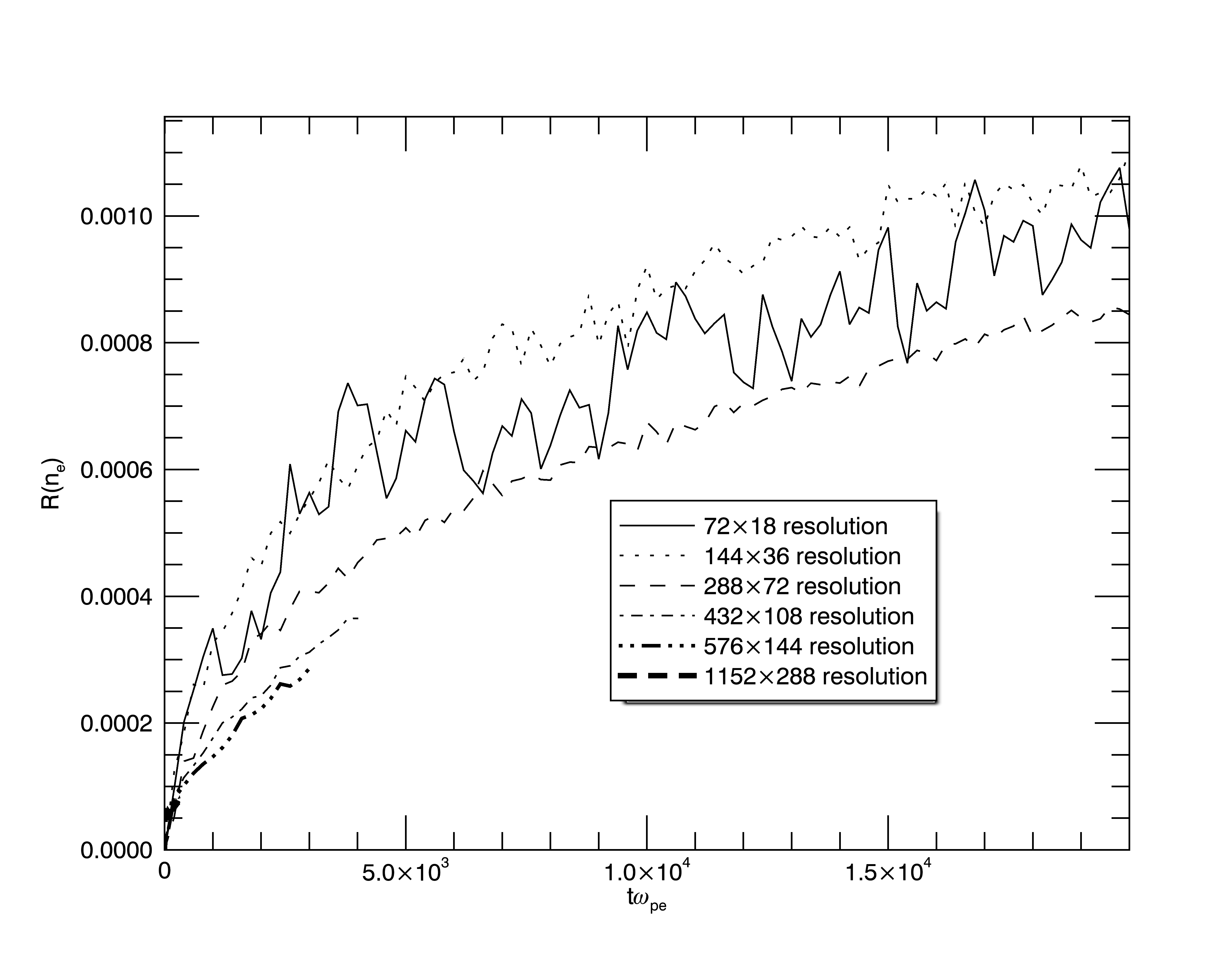}
\includegraphics[width=0.5\textwidth]{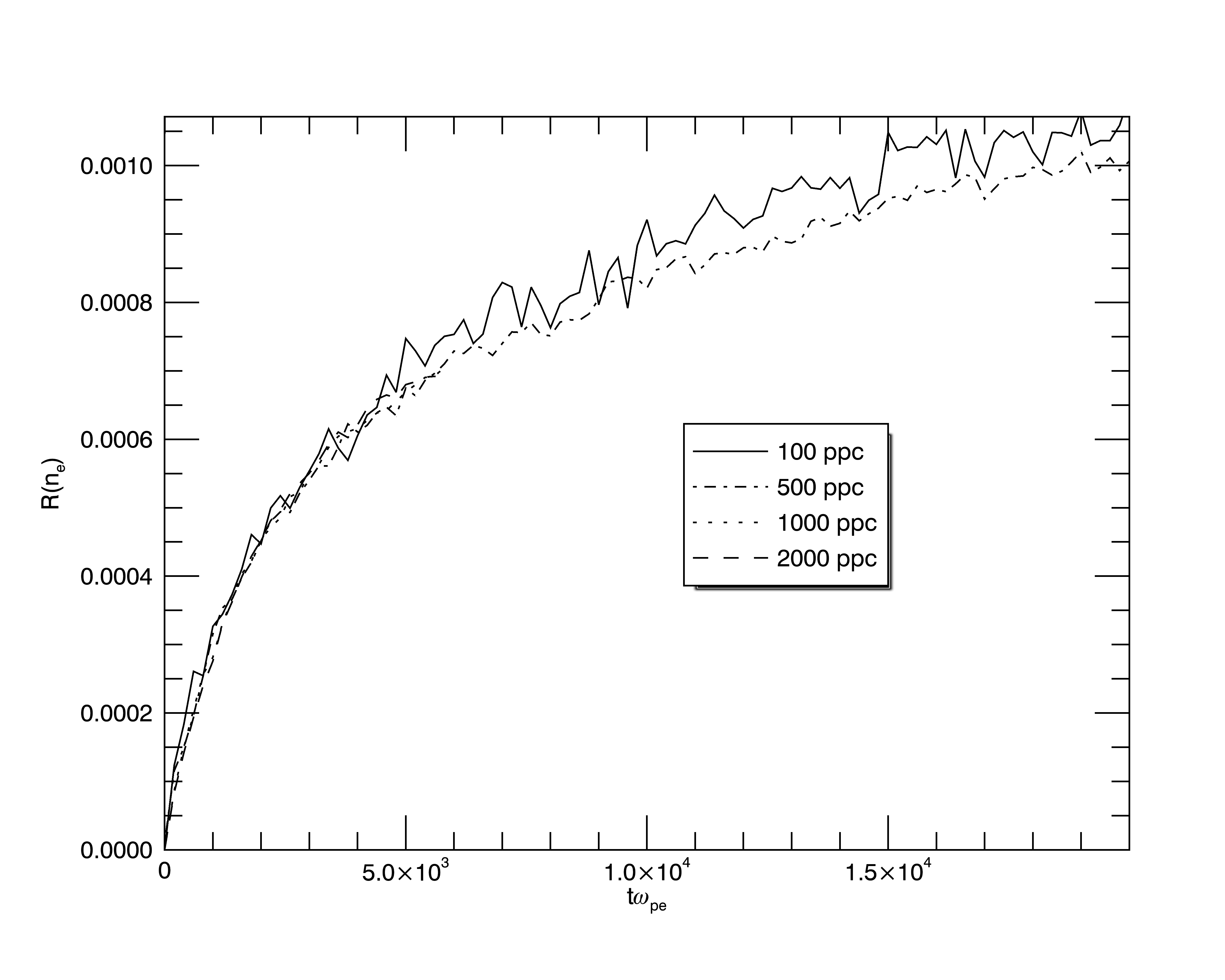}
\caption{ Top, mid and bottom panels 
show time evolution of $R(n_e(t))$ according to Eq.(1).
The aim of this figure is to
explore convergence of the numerical
results as (i) domain size in top panel, (ii) grid resolution in mid panel
and (iii) PPC in bottom panel are varied, accordingly.}
\label{f3}
\end{figure}

Before presenting main results, it is important to make sure that the 
simulation results are not very sensitive with respect to the numerical 
parameters of the EPOCH model. This means that by altering (i) the computational 
domain size, (ii) the grid density, and (iii) the number of particles per 
cell (PPC) the obtained solutions remain reasonably unchanged.

We define the following quantity that is a physical measure of free charge 
creation by collisional ionization
\begin{equation}
R(n_e(t))=\frac{\int_{0}^{L_x} \int_{0}^{L_y} (n_e(x,y,t)-n_e(x,y,0)) dxdy  }
{\sqrt{L_x L_y} \times \int_{0}^{L_x} \int_{0}^{L_y} n_e(x,y,0)dxdy }, 
\end{equation}
where $n_e(x,y,t)$ is number density of electrons, $L_x$ and $L_y$ are 
grid lengths in x- and y- directions.

It can be noted that because of the normalisation by the initial (at $t=0$) 
number density of electrons, effectively, $R(n_e(t))$ gives percentage of 
electrons scaled by a scaling factor of $1/\sqrt{L_x L_y}$ to compare 
solutions obtained for different ensemble sizes corresponding to 
different numbers of identical computational cells. The scaling 
factor comes from considering the collision of particles in cells as a 
random process in terms of the interaction between different cells of the 
computational domain similar to the classical diffusion as discussed in 
the end of this sub-section.

For the numerical parameter sensitivity study, the case of $C60:C12$ 
fraction of $50:50$ is selected. For the numerical integration in Eq.(1) 
an Interactive Data Language's (IDL's) built-in function is used ({INT\_TABULATED}). 
This function integrates a tabulated set of data $\{ X_i , F_i \}$ on the closed 
interval $\rm{[MIN(X) , MAX(X)]}$, 
using a five-point Newton-Cotes integration formula. 
The implementation is based on introducing of an auxiliary array in 
EPOCH which contains y-array with x-values integrated out. This is 
followed by integration of the y-dependence in order to obtain a 
single value of $R(n_e(t))$ at a given solution time $t$.

The top, mid and bottom panels of Fig. \ref{f3} examine sensitivity 
of the numerical solution when gradually changing (i) the domain size, (ii) 
the grid resolution, and (iii) the PPC number.
In the top panel of the figure, the grid unit is $\Delta=4 \lambda_D$ and 
grid size is increasing by an appropriate factor e.g. $384 \times 96$ for solid line.
In the mid panel, the domain size remains the same, for example, our standard 
grid resolution $288 \times 72$ has $\Delta=2 \lambda_D$
while $432 \times 108$ has $\Delta=(3/4) \lambda_D$, and commensurately 
$576 \times 144$ has $\Delta=1 \lambda_D$.

The bottom panel corresponds to the grid resolution of $144 \times 36$ 
with $\Delta=4 \lambda_D$ and varied PPC. This lower spatial resolution enabled 
us to access large PPC values while keeping the simulation cost feasible.
It can also be noted that in Fig. \ref{f3} not all lines go up to the final 
dimensionless simulation time, this is because all numerical runs have been 
limited by 10 day (240 hour) wall-time. Typical numerical run utilized 
circa 144 processing cores connected with Infiniband Interconnect.
In solutions presented on the mid and bottom panels in Fig. \ref{f3} 
the factor of $1/\sqrt{L_x L_y}$ from Eq.(1) correspond to different 
grid densities and varied numbers of particles per cell are. The solutions 
for four highest grid densities from the $288 \times 72$ resolution and 
for all PPC numbers are in a good agreement with one another.
The top fig. \ref{f3} shows that solutions for different domain sizes 
are in a reasonable agreement with the theoretical scaling depending 
on the statistical ensemble size. 

All-in-all this confirms that the suggested simulation results are 
reasonably non-sensitive to the numerical parameters of the EPOCH 
model for the parameter range of interest. In all cases, the free 
charge created by collisional ionization as a function of time has the 
same functional behaviour which can be explained by a simple analytical 
linear model as discussed in the end of this sub-section.

\begin{figure}
\centering
\includegraphics[width=0.5\textwidth]{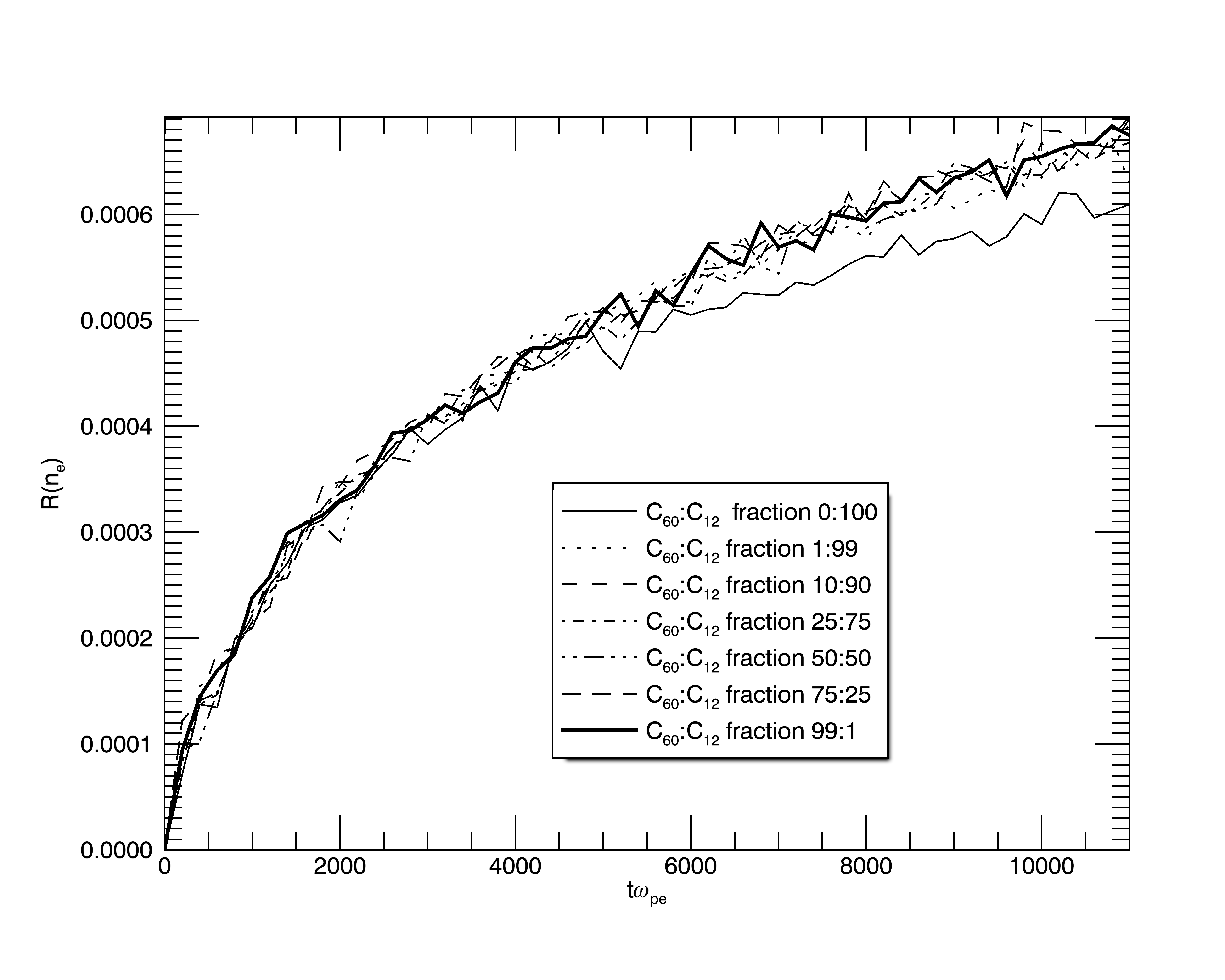}
\includegraphics[width=0.5\textwidth]{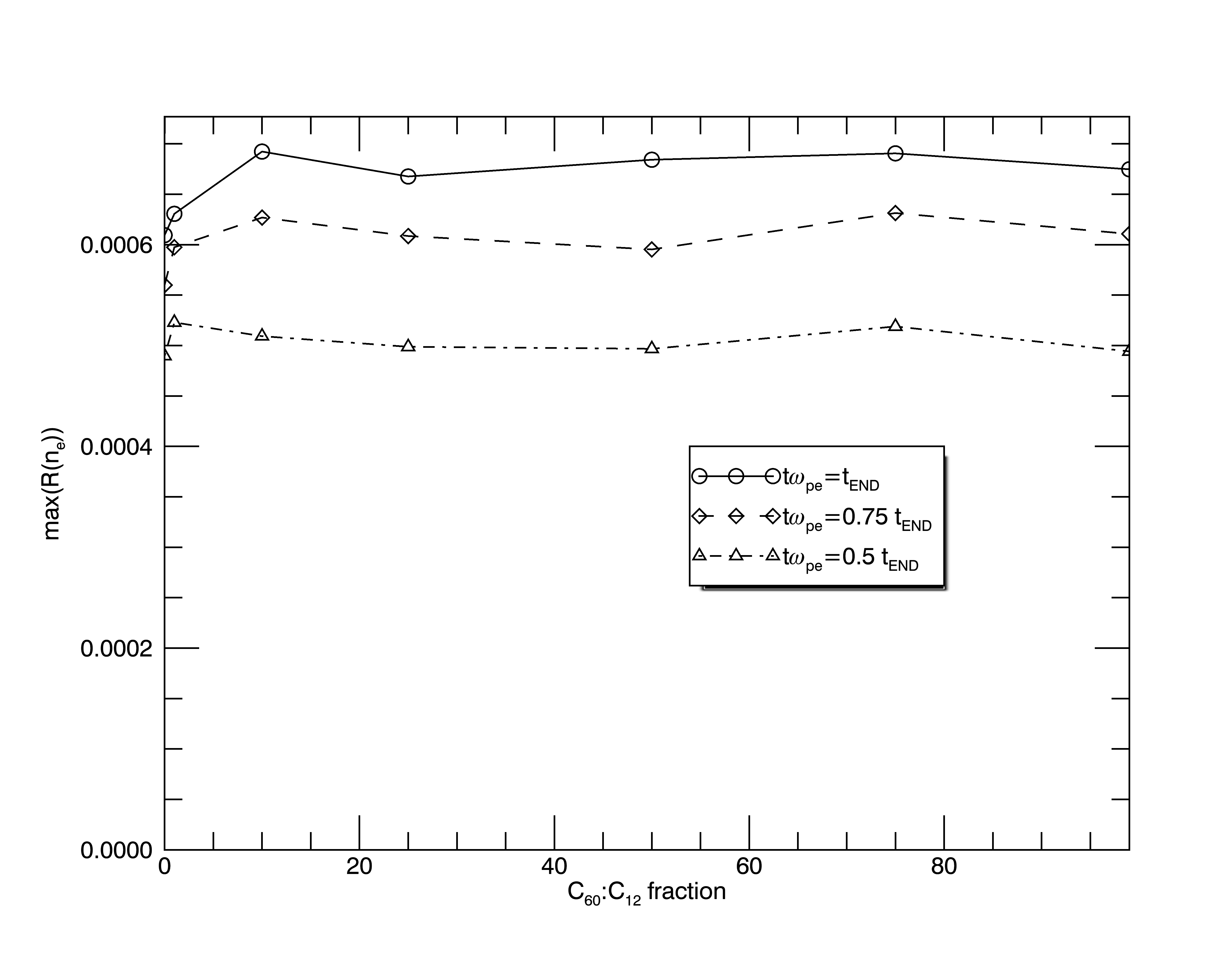}
\caption{ Top panel shows time evolution of $R(n_e(t))$, according to Eq.(1), 
for the different fractions: thin solid line is for 
$C60:C12$ fraction $0:100$, 
dotted line for $C60:C12$ fraction $1:99$, and so on, see panel inset
for details,
until thick solid line corresponding to $C60:C12$ fraction down to $99:1$.
Bottom panel shows 
$R(n_e(t=t_{END}))$, i.e.  $R(n_e(t)$ at the final simulation time with 
solid line with open circles,
$R(n_e(t=0.75 t_{END}))$ dashed line with open diamonds,
$R(n_e(t=0.5 t_{END}))$ dash-dotted line with open triangles.}
\label{f4}
\end{figure}

In Fig. \ref{f4} we explore the effect of different $C60:C12$ fractions
on free electron production. The fractions are  as follows: $0:100$, $1:99$, 
$10:90$, $25:75$, $50:50$, $75:25$, $99:1$.
Top panel shows time evolution of $R(n_e(t))$ for these different fractions. 
One important aspect immediately seen in this panel is that the absence of 
C60 yields smallest possible free electron production by collisional ionization.
Even adding 1\% of C60 markedly changes the situation.
Bottom panel shows 
$R(n_e(t=t_{END}))$ solid line with open circles,
$R(n_e(t=0.75 t_{END}))$ dashed line with open diamonds,
$R(n_e(t=0.5 t_{END}))$ dash-dotted line with open triangles.

From the above results, a local optimum for free electron production by 
collisional ionization (i.e. a most efficient discharge condition) occurs 
for $C60:C12$ fraction of $10:90$. There is a second optimum in the flat 
part of the distribution which corresponds to $C60:C12$ fraction of about $80:20$. 
The bottom panel shows that the two optima are fairly persistent in the data 
once the simulation time becomes greater than $t=0.5 t_{END}$.

Interestingly, the optimum corresponding to 10\% of C60 is in a broad agreement 
with the preliminary experimental results of the 
Engineering Company Eco-Ardens who explored 
the efficiency of tribo-electric plasma generation in their gasifier for 
relatively small (less than 50\%) fractions of C60 in the gas mixture.

\begin{figure} 
\centering
\includegraphics[width=0.5\textwidth]{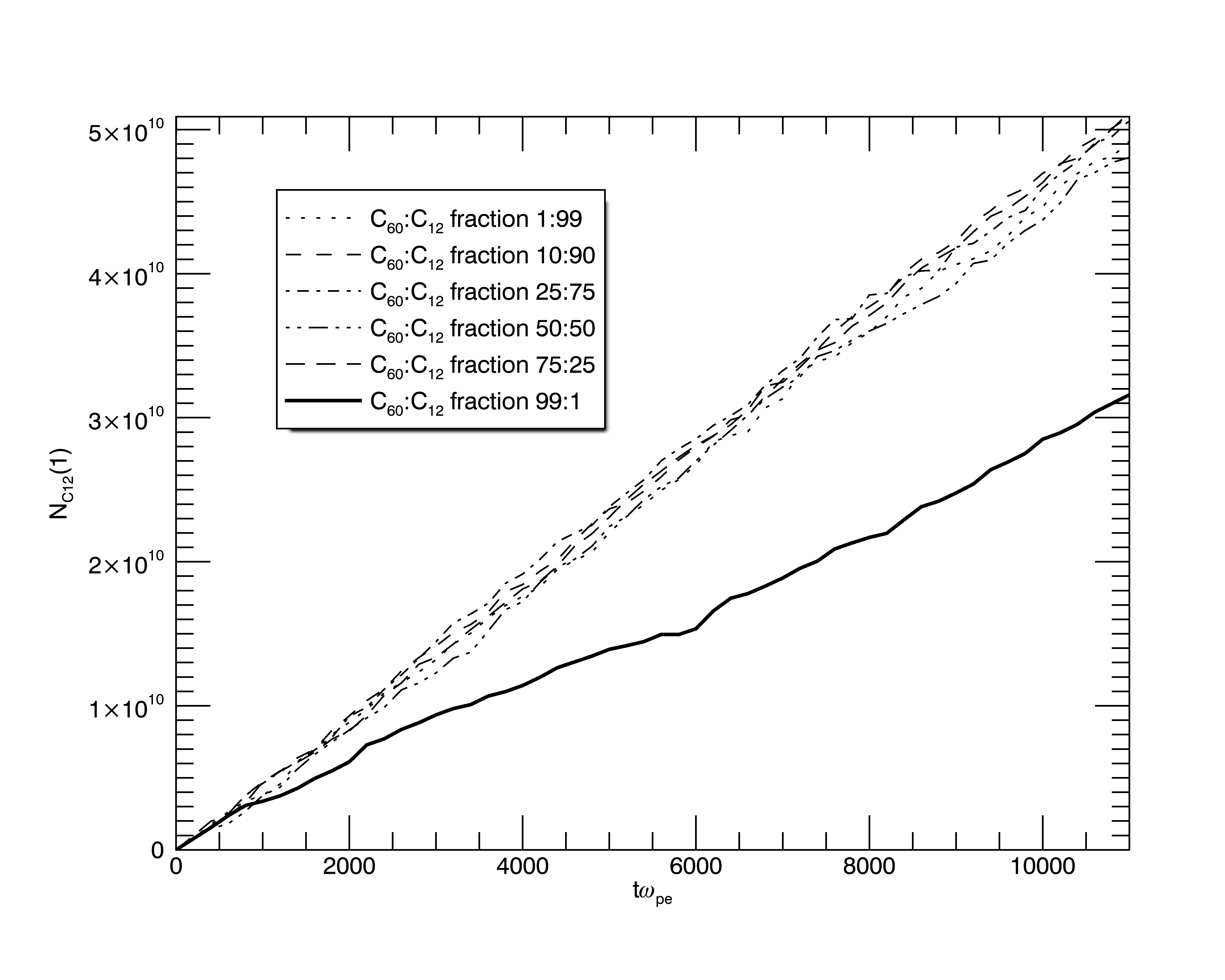}
\includegraphics[width=0.5\textwidth]{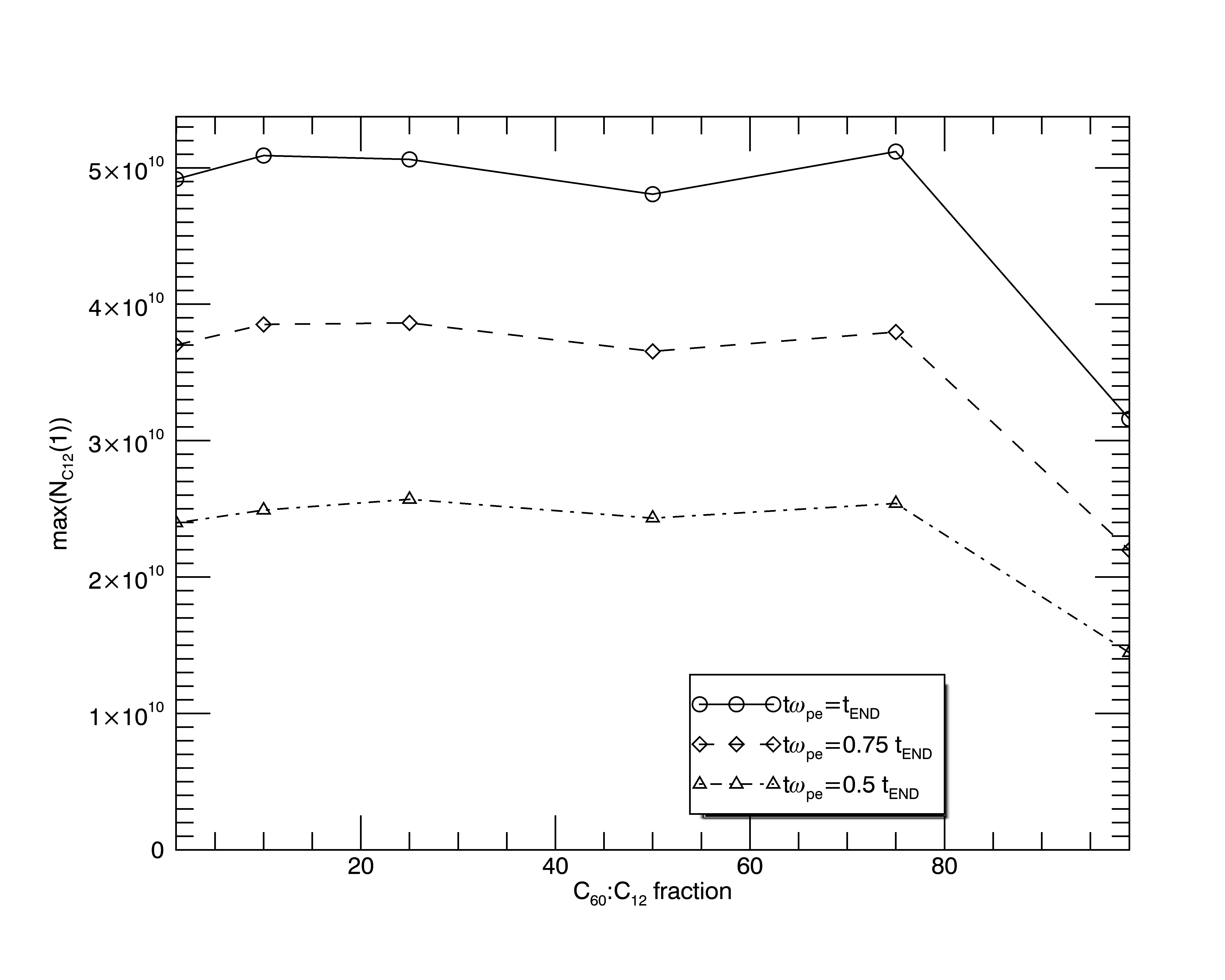}
\caption{ In the top panel we plot $N_{\rm C12}(1)(t)$, according to Eq.(2), 
for different $C60:C12$ fractions -- see panel inset
for details. 
In the bottom panel we plot  
$N_{\rm C12}(1)(t=t_{END})$ solid line with open circles,
$N_{\rm C12}(1)(t=0.75t_{END})$ dashed line with open diamonds,
$N_{\rm C12}(1)(t=0.5t_{END})$ dash-dotted line with open triangles.}
\label{f5}
\end{figure}

Fig. \ref{f5} shows the effect of different $C60:C12$ fractions
on C12+ (singly ionized C12) production by collisional ionization.
In contrast to free electrons which are always present because of the 
background plasma field, there are no C12+ at $t=0$.

Hence, instead of a definition similar to Eq.(1)  we quantify the C12+  production using
\begin{equation}
N_{\rm C12}(1)(t)=\frac{\int_{0}^{L_x} \int_{0}^{L_y} n_{\rm C12}(1)(x,y,t) dxdy  }{\sqrt{L_x L_y}  },  
\end{equation}
where $n_{\rm C12}(1)(x,y,t)$ is number density of C12+.

The top panel of Fig. \ref{f5}, shows $N_{\rm C12}(1)(t)$ for 
different $C60:C12$ fractions. This quantity increases approximately 
linearly in time and different $C60:C12$ fractions have different growth rates.
The difference of growth rates can be more readily seen in the 
bottom panel of Fig. \ref{f5} where we plot  
$N_{\rm C12}(1)(t=t_{END})$ solid line with open circles,
$N_{\rm C12}(1)(t=0.75t_{END})$ dashed line with open diamonds,
$N_{\rm C12}(1)(t=0.5t_{END})$ dash-dotted line with open triangles.

Similar to the free charge distribution (comp. with Fig. \ref{f4}), 
there are the same two optima corresponding to $C60:C12$ fractions of 
$10:90$ and $80:20$. However, the first maximum ($C60:C12$ of $10:90$) 
is rather flat while the second peak is more prominent. The bottom 
fig. \ref{f4} shows that the broad and the sharp maxima are notable 
in the distribution for all solution times.

It should be noted that the values attained by $N_{\rm C12}(1)$ are much smaller 
than the number density of electrons in the plasma $n_e$ as well as the number of 
free electrons generated by collisions.
For example, by making the following substitution 
$n_{\rm C12}(1)(x,y,t) \to n_e(x,y,t)$ in Eq.(1) for $C60:C12$ fraction of $10:90$ 
and taking $t=t_{END}$, it follows that $N_{\rm n_e}(t=t_{END})=1.98\times 10^{14}$ 
which gives the number of generated free electrons. By comparison with the maximum 
value of $N_{\rm C12}$
(shown in the top panel of fig. \ref{f5}), the number of free 
electrons generated by collisions is three orders of magnitude larger.

\begin{figure}
\centering
\includegraphics[width=0.5\textwidth]{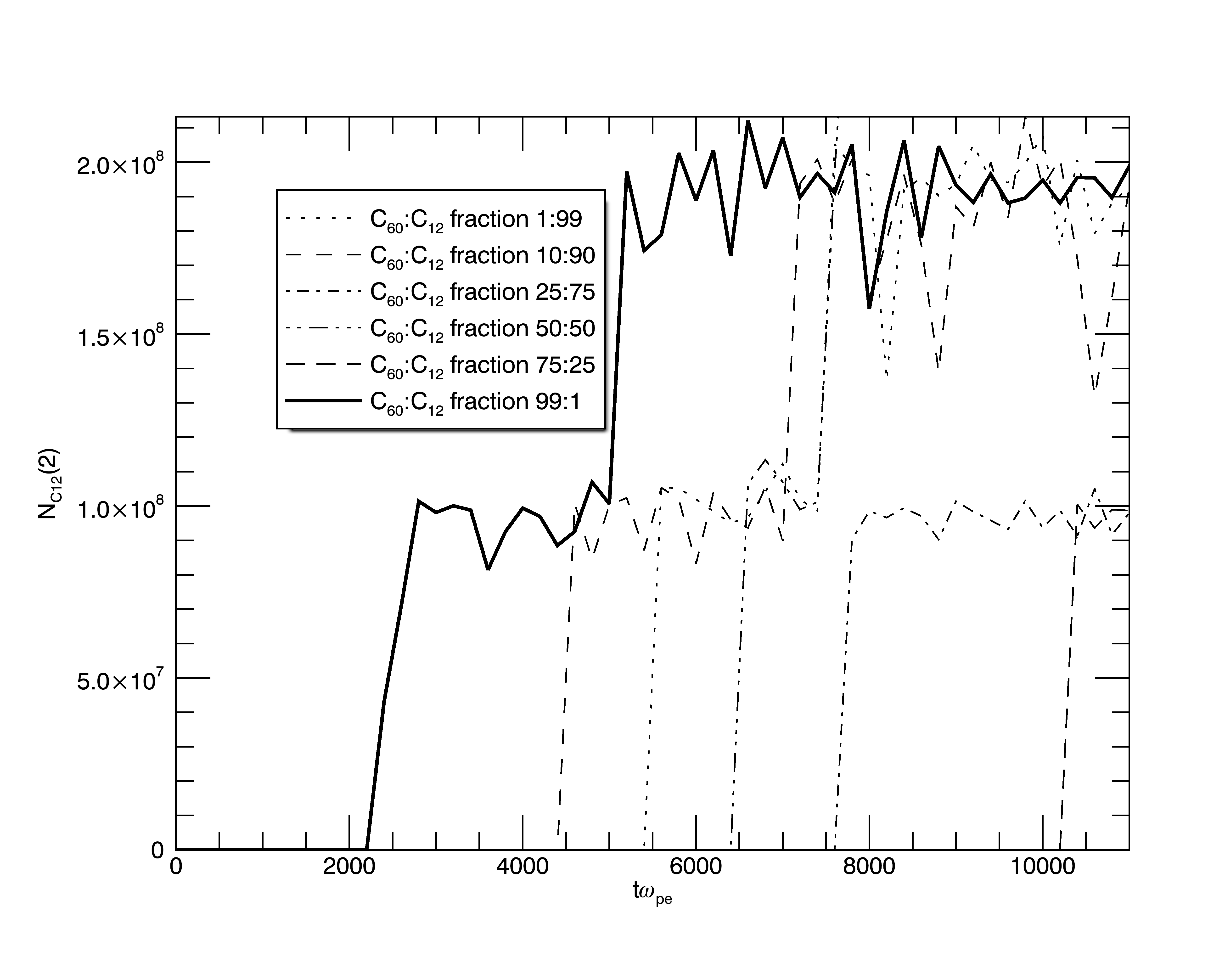}
\includegraphics[width=0.5\textwidth]{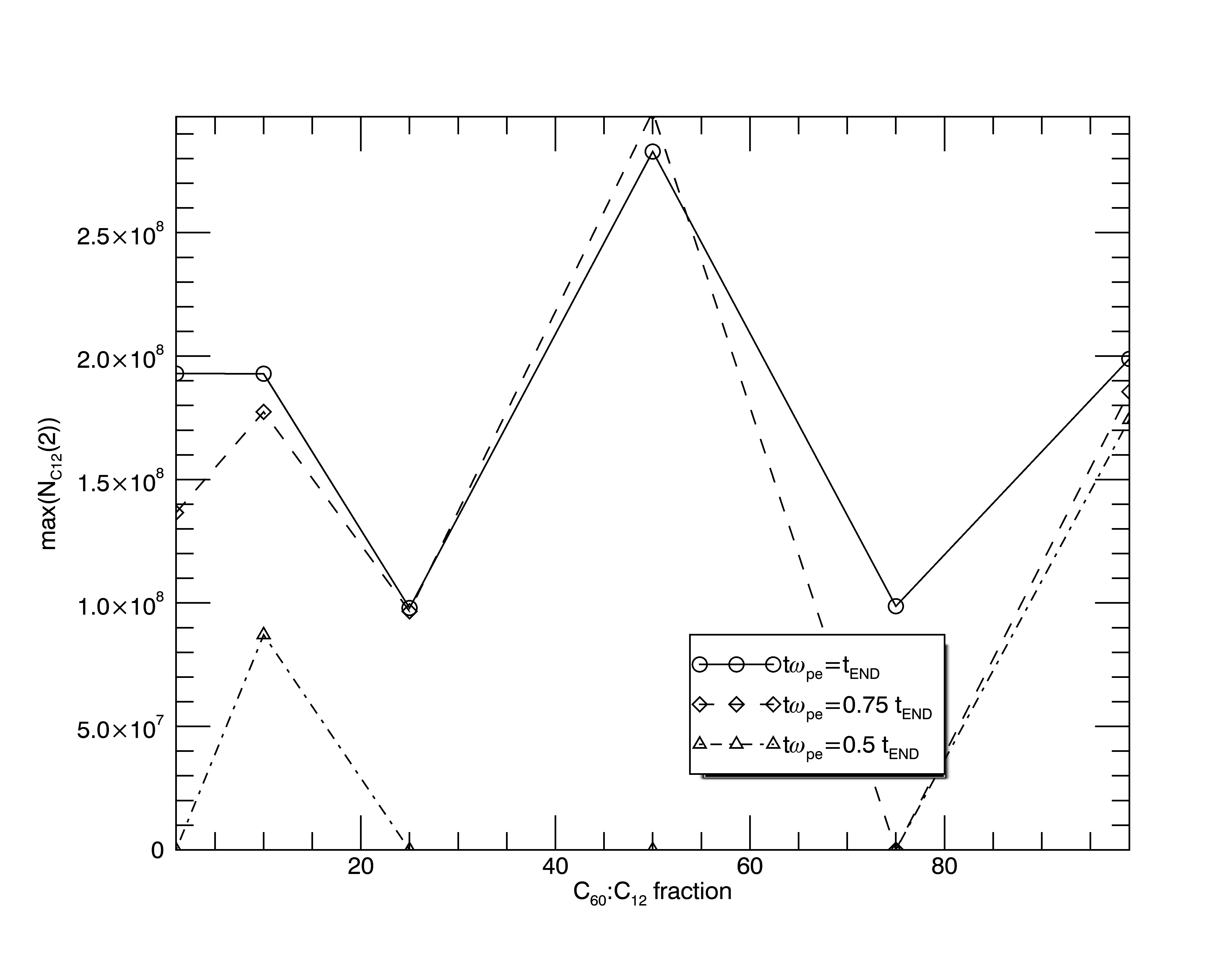}
\caption{ Time evolution of
$N_{\rm C12}(2)(t)$, defined  by Eq.(2), as in Fig.5, but  replacing 
$n_{\rm C12}(1)(x,y,t)$ now with $n_{\rm C12}(2)(x,y,t)$, 
the latter being  number density of C12++.
Here, as in Fig.5, we study of the effect of different $C60:C12$ fractions
on C12++ (doubly ionized C12) production by collisional ionization
-- see panel insets
for details. }
\label{f6}
\end{figure}

Fig. \ref{f6} demonstrates the effect of different $C60:C12$ fractions on 
C12++ (doubly ionized C12) production by collisional ionization.
$N_{\rm C12}(2)(t)$ is defined as by Eq.(2) but simply replacing 
$n_{\rm C12}(1)(x,y,t)$ by $n_{\rm C12}(2)(x,y,t)$ with the latter being  number density of C12++.
Two observations follow from top panel Fig. \ref{f6}:
(i) there is no longer monotonous increase of C12++ production by 
collisional ionization. Instead, the process proceeds in jumps. For example, 
for the case of $C60:C12$ fraction of $10:90$, represented by a dashed line, 
there are two jumps at $t \omega_{pe}=4500$ and $6500$.
(ii) the obtained number densities of C12++ are further three order of 
magnitude smaller in comparison with the singly ionized C12 case, e.g. 
$N_{\rm C12}(2)(t=t_{END})=2\times 10^{8}$ for most cases.
(iii) there are three peaks in the number density distribution of C12++ as a 
function of the $C60:C12$ fraction. In addition to $C60:C12$ fraction of $10:90$, 
two new maxima include $C60:C12$ fraction of $50:50$ and another peak at high 
fullerene concentrations tending to $100:0$.

\begin{figure} 
\centering
\includegraphics[width=0.5\textwidth]{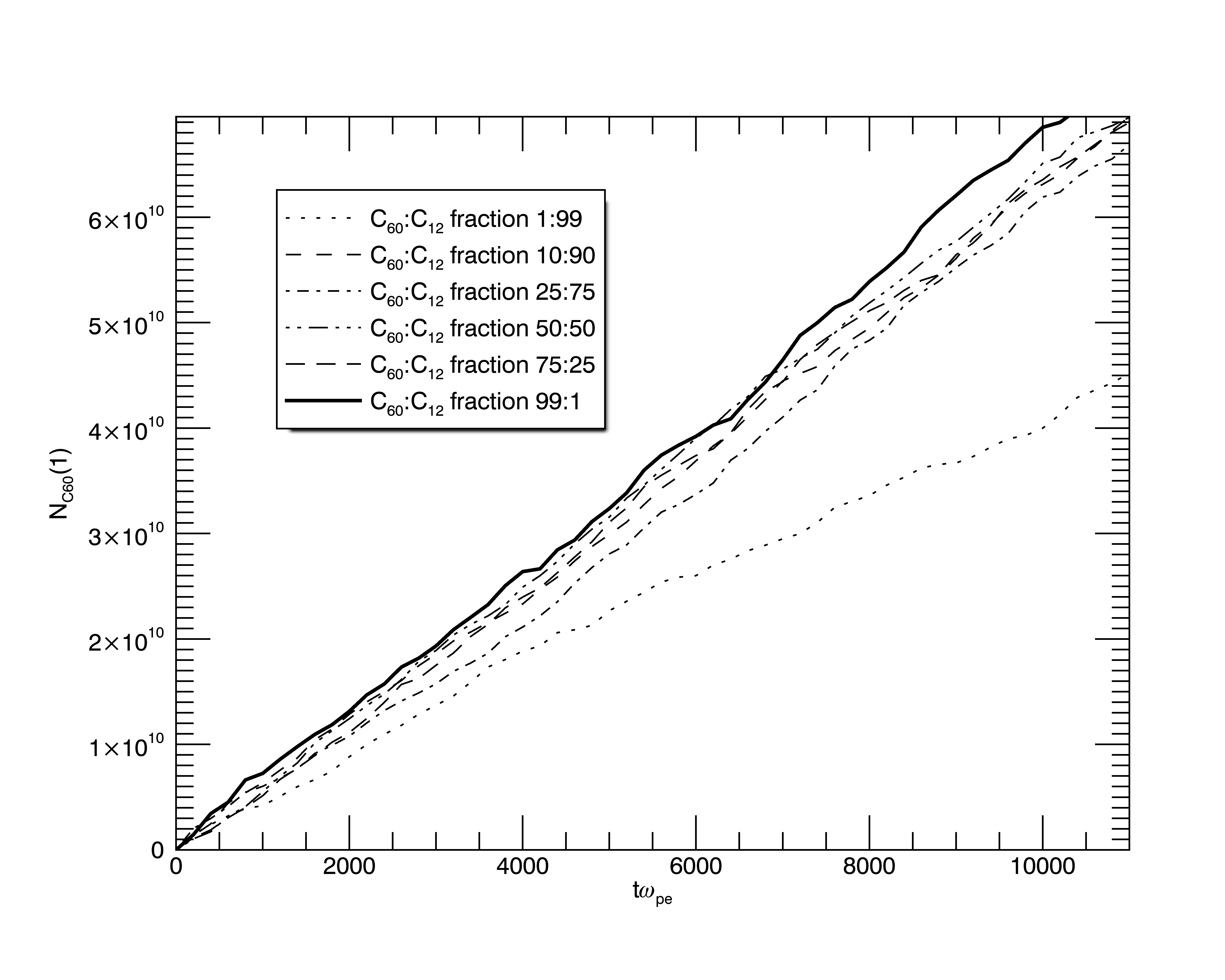}
\includegraphics[width=0.5\textwidth]{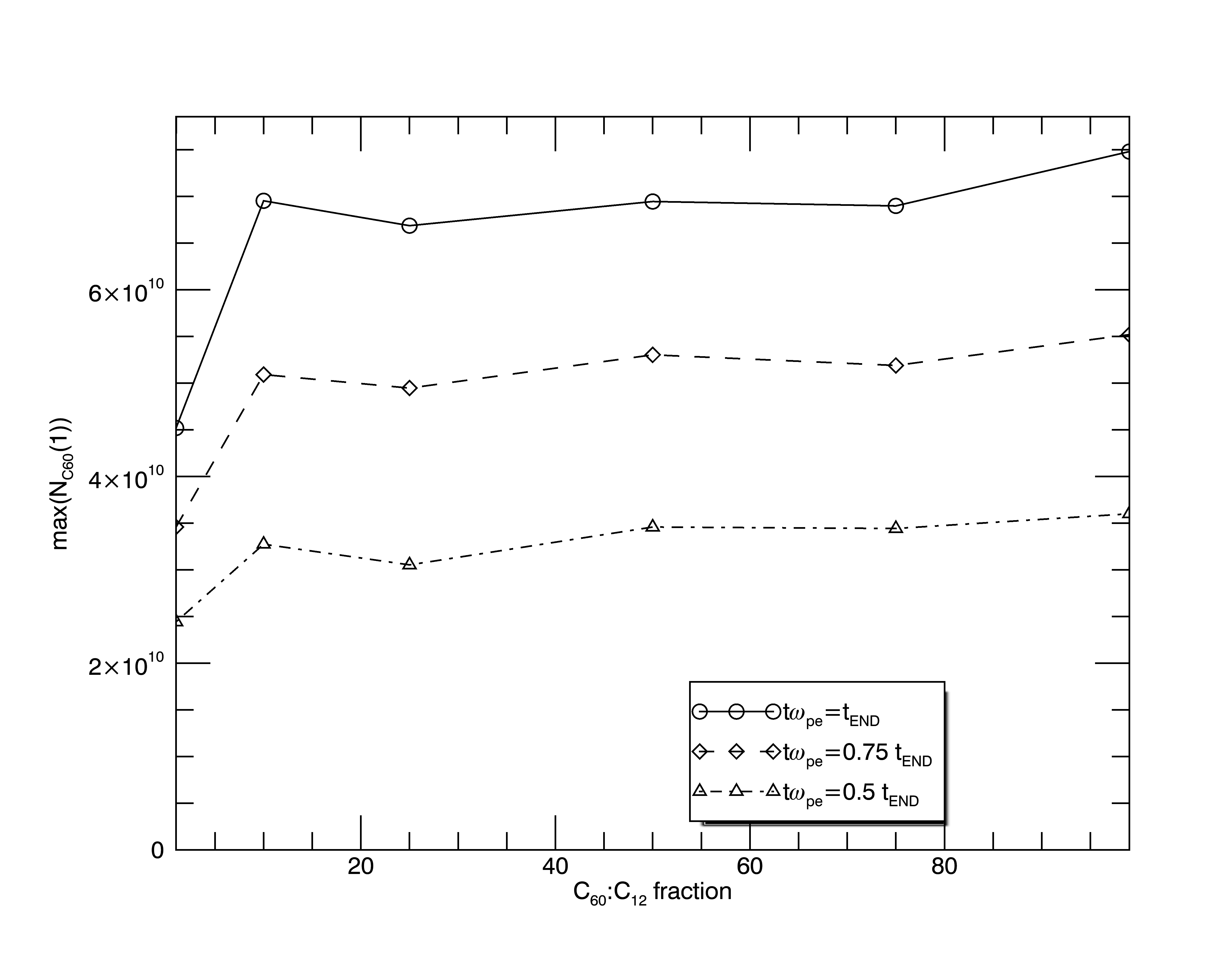}
\caption{ As in Fig. 5, but now investigating the effect of different $C60:C12$ fractions
on C60+ (singly ionized C60, i.e. fullerene) production by collisional ionization. }
\label{f7}
\end{figure}

Fig. \ref{f7} shows the effect of different $C60:C12$ fractions
on C60+ (singly ionized C60, i.e. fullerene) production by collisional ionization.
Broadly speaking, the behaviour of C60+ is similar 
to that on of C12+ shown in Fig. \ref{f5}, except
for:
(i) the number density values of C60+ is about 30\% larger in comparison with that of C12+ 
and
(ii) the peak at $C60:C12$ fraction of $10:90$ is more clearly pronounced and the broad 
peak at high fractions of $C60:C12$ moves to $100:0$.

To conclude this section, an analytical ionization model is considered where the 
C12 and C60 species are lumped together as the two-species (ionised and non-ionised) 
of a particle gas immersed in plasma.
Following ref.\cite{kulk}, by introducing constant ionization and recombination parameters
$\alpha = const > 0$ and $\beta = const > 0$,
assuming that the number of electrons per unit volume is approximately constant, $n_e =
const > 0$ (i.e. the change of the electron number density due to ionization is much smaller in
comparison with the original electron number density in plasma,
 $n_e(t)/n_e(0) \approx 1$),
and denoting the numbers of ionized and non-ionized particles by $n_1$ and $n_2$, respectively, so
that $n_1 + n_2 = n_0 = const > 0$, where $n_0$ is the total number of particles per unit volume
that is fixed constant, the evolutionary equation for the ionized particles is given by:
\begin{equation}
\frac{\partial n_1}{\partial t} = \alpha n_e n_2  -\beta n_e n_1 .  
\end{equation}
Equation (3) is solved in a periodic spatial domain
$0 \leq x \leq L_x$, and $0 \leq y \leq L_y$
with the initial condition:
$n_1 (x, y, 0) = 0$
and under the constraint that
$n_1 + n_2 = n_0 $.

From integration of Eq.(3) over the control volume $V =\int_{0}^{L_x} \int_{0}^{L_y}  dxdy$ one obtains:
\begin{equation}
\frac{d \langle n_1\rangle}{d t} = \alpha N_e \langle n_2 \rangle -\beta N_e \langle n_1 \rangle,  
\end{equation}
where $\langle f \rangle=\int_{0}^{L_x} \int_{0}^{L_y} f dxdy$.
The constraint after averaging yields $\langle n_2 \rangle=\langle n_0  \rangle - \langle n_1 \rangle $, 
which is then substituted into Eq. (4) to obtain
\begin{equation}
\frac{d \langle n_1\rangle}{d t} = \alpha N_e \langle n_0 \rangle -(\beta+\alpha) N_e \langle n_1 \rangle,  
\end{equation}
where $N_e$ is the electron number in the considered control volume $V $assuming that the 
non-linear process in the bigger volume leads to an appropriate renormalisation of the coefficients
$\alpha$ and $\beta$.
By introducing new notations
$\langle n_i \rangle = \langle n_1 \rangle / \langle n_0 \rangle$, 
$a=(\alpha+\beta)N_e$ and $b=\alpha/(\alpha+\beta)$
Eq.(5) simplifies to
\begin{equation}
\frac{d \langle n_i\rangle}{d t} = -a(\langle n_i \rangle -b).  
\end{equation}
Using the initial condition, the solution for the averaged particle number in the control volume $V$ is
\begin{equation}
 \langle n_i\rangle(t) = b \left(1-\exp(-a t) \right).  
\end{equation}
The above analytical solution can be compared with predictions of the 
relative change in electron number density computed using EPOCH code,
$R(n_e ) =\left(\langle n_e(t) \rangle -\langle n_e(0) \rangle \right) /\langle n_e(0) \rangle$,
which essentially coincides with our definition for $R(n_e )$ from Eq.(1).
First of all, note that the number of ionized particles scales with the number of new electrons
generated such that
\begin{equation}
\langle n_i\rangle = C \left(\langle n_e(t) \rangle -
\langle n_e(0) \rangle \right) /\langle n_e(0) \rangle,  
\end{equation}
where $C = const > 0$.
Hence,
$\ln(  \langle n_i\rangle ) = \ln(  R(n_e) ) + \ln(C)$ and, using Eq.(7)
\begin{equation}
\frac{d \ln(  R(n_e) )}{d t} = \frac{d \ln(  \langle n_i\rangle )}{d t}=
\frac{a}{1-\exp(-at)}.  
\end{equation}

For initial times $at \ll 1$, Eq.(9) can be further simplified using the Taylor expansion that leads
to the asymptotic solution as follows
\begin{equation}
\frac{d \ln(  R(n_e) )}{d t} \simeq 
\frac{1}{t}.  
\end{equation}

\begin{figure}
\centering 
\includegraphics[width=0.5\textwidth]{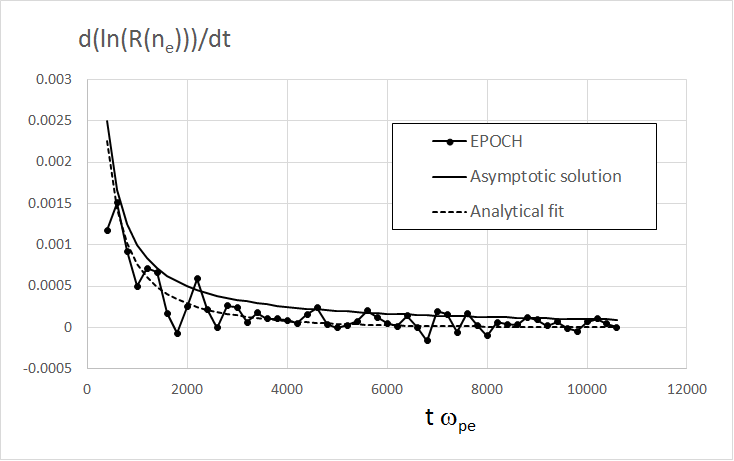}
\caption{ Solution of the homogeneous particle ionization problem: comparison of the
EPOCH solution (solid line with dots) with the analytical solutions, where solid line shows
 asymptotic solution according to Eq.(10) and 
dashed line is analytical solution, according to Eq.(9), 
in which the value of parameter $a$ has
been adjusted to obtain the best fit with the EPOCH solution. }
\label{f8}
\end{figure}

Fig. \ref{f8} shows comparison of the EPOCH solution with the analytical solution 
Eq. (9) and the asymptotic solution Eq. (10). In the case of the analytical model 
Eq. (9), the value of parameter $a$ has
been adjusted to obtain the best fit with the EPOCH solution.

It can be noted that, despite some noise present in the EPOCH data due to the
 numerical differentiation, the analytical solutions based on 
 Eqs.(9) and especially (10) are in a good agreement with the numerical solution.

By recalling the need to explain the scale-factor  of $1/\sqrt{L_x L_y}$ 
from Eq.(1) mentioned when we discussing top panel Fig. \ref{f3}, we next explore the 
effect of the periodic boundary condition for comparison of the simulation results in different domain sizes.

Let us consider the solution of 
Eq. (5) in a large domain, $V^{N,M} = \int_{0}^{N\times L} \int_{0}^{M\times H} dxdy$, 
where $N, M >1$ are the total number of grid cells and the x- and the y- direction, respectively. 
The control volume $V$, considered in the previous analysis can be treated 
as subset of the large domain. The goal is to compare the particle number 
density solution, Eq. (7), obtained in the domain
$V$ and the same averaged over the larger domain $V^{N,M}$.
To proceed, the large domain is broken down in several over non-overlapping 
sub-volumes $V^{k,l} =\int_{k \times L}^{(k+1)\times L} \int_{l\times H}^{(l+1)\times H}  dxdy$, 
where $1 < k < N$ and $1 < l < M$. Each of these sub-volumes $V^{k,l}$
 is equal to $V$ but, in comparison with the single volume case, the particle collision
processes in separate sub-volumes are largely uncorrelated with one another.
The particle numbers averaged over each sub-volume $V^{k,l}$ satisfy to Eq(6). 
Each of these quantities can be treated as random variables, whose 
evolutionary equations can be treated by
analogy with the Langevin diffusion
\begin{equation}
\frac{d n_i }{d t} = -a n_i+ \{R\}, 
\end{equation}
where $\{R\} = ab$ is the generation term that can be interpreted as a random force and brackets
of the volume averaging in the particle number variable are omitted. Here $a=(\alpha+\beta)N_e$, where
$N_e$ is the electron number corresponding to the large domain ensemble.
In accordance with the well-known solution of the Langevin equation \cite{lan}, 
the variance of the ensemble averaged number of the particles grows as
$$
\langle n_i^2\rangle (t) =
\left[\langle n_i^2\rangle (0)-Amp(R)^2/(2a) \right] \exp(-2at)
$$
\begin{equation}
+Amp(R)^2/(2a), 
\end{equation}
where $Amp(R) = ab$.
Hence,
$$
\left(\langle n_i(t) \rangle -\langle n_i(0) \rangle \right) \simeq
\sqrt{\langle (n_i^2(t) \rangle}=
$$
\begin{equation}
\sqrt{
\left[\langle n_i^2\rangle (0)-ab^2/2 \right] \exp(-2at)
+ab^2/2}. 
\end{equation}
At equilibrium, $\left(\langle n_i(t) \rangle -\langle n_i(0) \rangle \right)=\sqrt{a/2}b=
\sqrt{N_e/2}\frac{\alpha}{(\alpha+\beta)}$
 and the quantity
 $\left(\langle n_i(t) \rangle -\langle n_i(0) \rangle \right)/\sqrt{N_e}=\sqrt{\frac{\alpha}{2(\alpha+\beta)}}$
 should be independent
of the size of the considered system, $\simeq NM$.
Using Eq.(8) this leads to the following scaling of the simulation results for different size
periodic domains:
\begin{equation}
R(n_e )/\sqrt{N_e} =const 
\end{equation}
The top panel Fig. \ref{f3} shows the simulation results for
different domain sizes. 
It can be noted that the revealed dependency of the ionized particle solution on the domain size
is similar to the so-called "shot noise" effect reported in the start-up laser problems \cite{shot}.
Because (i) $N_e \propto NM \propto L_x L_y$ and (ii) all lines
for the different domain sizes in top panel of Fig. \ref{f3} are tolerably close
to each other, hence scale-factor  of $1/\sqrt{L_x L_y}$ from Eq.(1) is justified based on
our Langevin equation solution.

\subsection{Particle interaction with including the non-homogenous wall condition}

The EPOCH results describing how surface roughness 
affects free charge generation by collisional ionization 
are presented next.
Instead of including actual material rough walls in the 
simulation, computationally, it is much easier to impose periodic 
electric field on the domain boundary. Indeed, rough surfaces 
alter electric field in the vicinity of the solid boundaries and 
the imposition of a non-uniform electric field boundary condition 
along with the periodic condition on particles is equivalent to 
considering a small internal volume of the particle domain at some 
distance away from the material walls. It can be reminded that 
enforcing of the periodic condition is important for consistency with the Maxwell's equations.

In EPOCH code, the boundary condition on the electric and 
magnetic field dynamics is implemented via a subroutine called 
fields.f90, see for details ref.\cite{ref16}. The following 
target electric fields at $y=y_{\rm min}=1$ are 
considered ($x$ is tangential to the wall boundary and $y$ is the normal direction):

(i) $E_y(x,t)=f(x,t)$
and
(ii) $E_x(x,t)=f(x,t)$,
where
$$
f(x,t) = E_0\times \biggl(
\exp(-(x-0.2 x_{\rm max})^6/(x_{\rm max}/15)^6)+
$$
$$
\exp(-(x-0.4 x_{\rm max})^6/(x_{\rm max}/15)^6)+
$$
$$
\exp(-(x-0.6 x_{\rm max})^6/(x_{\rm max}/15)^6)+
$$
$$
\exp(-(x-0.8 x_{\rm max})^6/(x_{\rm max}/15)^6)\biggr)
$$
\begin{equation}
\times \left[1.0-\exp\left(-t/(10/\omega_{pe})\right)\right], 
\end{equation}
and where $E_0=10^7$ Vm$^{-1}$.

The boundary condition at $y=y_{\rm min}=1$ is 
driven to the target field Eq(15) so that in about 
$t \omega_{pe}=10$ a steady state electric field with an amplitude of $E_0$ is 
reached. Such driving essentially imposes a comb-like electric field with 
four spikes at locations of $0.2, 0.4, 0.6, 0.8$ fraction of the 
computational domain size in the x-direction.

\begin{figure*} 
\includegraphics[width=\textwidth]{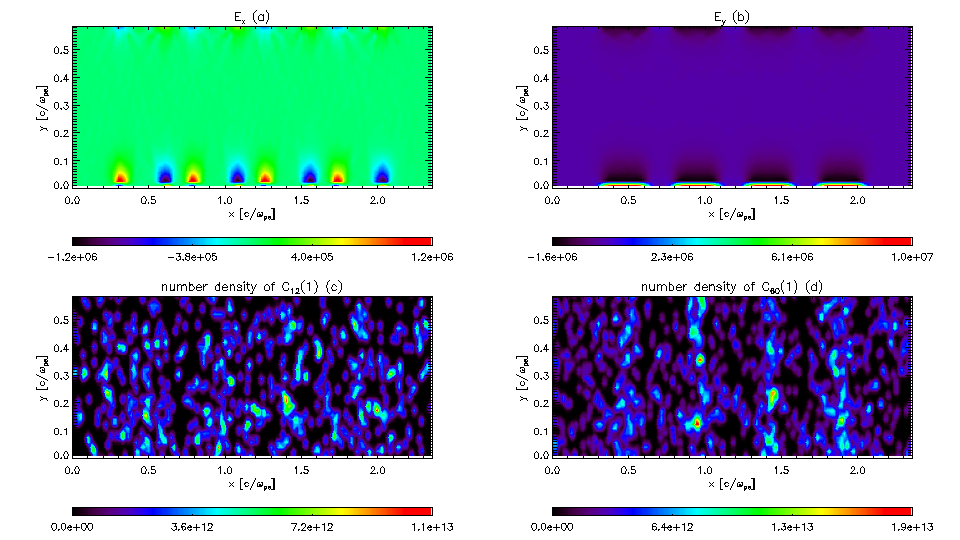}
\caption{ Top panels (a) and (b) show electric field x- and y- components,
respectively,
while bottom panels (c) and (d) should number densities of 
C12+  and C60+ 
at final simulation time $t \omega_{pe}=20000$, respectively.
The data is for driving the electric field component normal to the boundary. }
\label{f9}
\end{figure*}

Fig. \ref{f9} shows simulation results for the case of electric 
field component normal to the boundary.
The top panels demonstrate electric field x- and y- components, 
and bottom panels show number densities of C12+ (singly ionized C12) 
and C60+ (singly ionized C60) at final simulation time $t \omega_{pe}=20000$. 
The length scale units are based on the plasma frequency and the light speed. 

In the top panels of Fig. \ref{f9}, the electric field gradients are very 
localised and moderately penetrate in the domain interior.
The 'hot spots' which emerge in the two bottom panels of Fig. \ref{f9} 
represent charged ions of the relevant species. These species are relatively 
rare and more-or-less scattered over the whole domain.

\begin{figure*} 
\includegraphics[width=\textwidth]{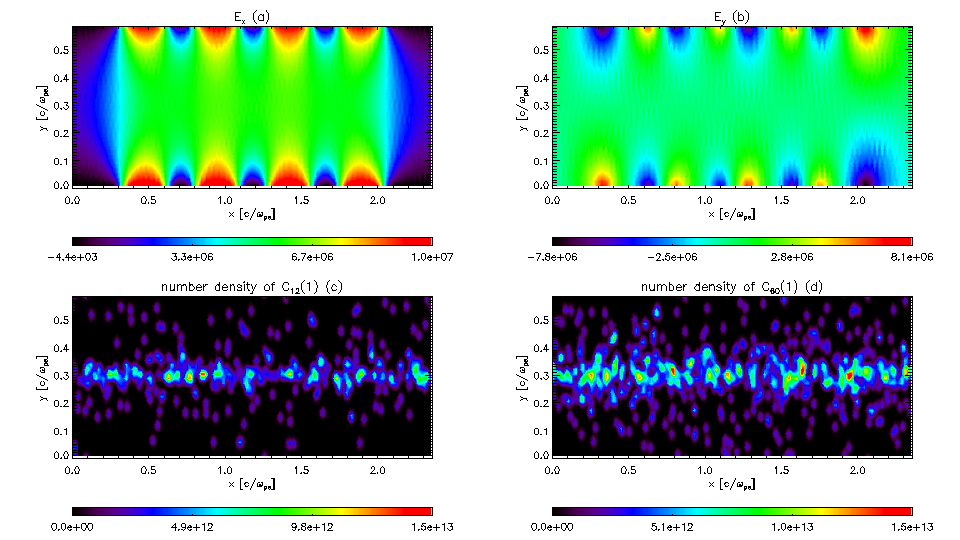}
\caption{The same as in Fig.\ref{f9} but for the case of the electric 
field tangential to the wall boundary}
\label{f10}
\end{figure*}
Fig. \ref{f10} shows simulation results for the case of tangential to 
x-direction electric field. The top panels show electric field x- and y- 
components, and bottom panels should number densities of C12+  and C60+ 
at final simulation time $t \omega_{pe}=20000$.
Two important observations from Fig. \ref{f10} include:
(i) $E_x$ now protrudes into the simulation domain much 
deeper than in the case of normal electric field driving and the 'flames' of the electric field gradient are much wider;
(ii) the 'hot spots' of C12+ and C60+ are clustered in the middle of the 
simulation domain at $y=y_{\rm max}/2$.

\begin{figure}
\centering 
\includegraphics[width=0.5\textwidth]{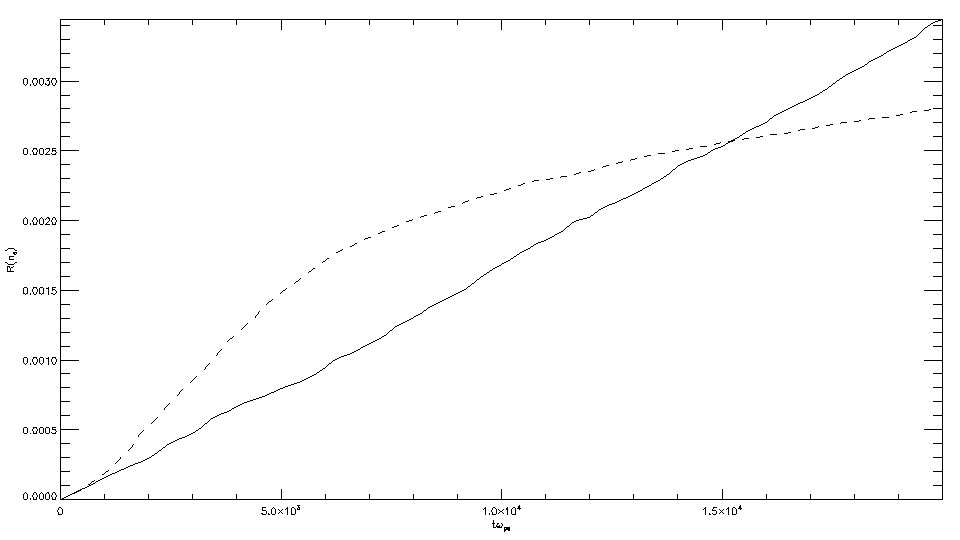}
\caption{$R(n_e)$ for the normal and tangential electric field driving cases.
Solid line is for the case of normal and dashed for the 
case of tangential electric field driving.}
\label{f11}
\end{figure}

Fig. \ref{f11} compares the time evolution of the 
previously defined relative integral charge, $R(n_e)$ that was generated 
in the case of the normal and the tangential electric field boundary condition.
Solid line is for the case of normal and dashed for the case of tangential electric field driving.
For the normal electric field case, the generated free charge, $R(n_e)$  
shows an approximately linear behavior without a sign of saturation. The suggests 
that the end simulation time has not been long enough to reach a quasi-steady 
state in this case.
However, in the case of the tangential electric field, the accumulated 
charge curve shows a sign of exponential stagnation towards a saturated 
state in accordance with the linear collision model discussed in the previous section.

Importantly, the charge localisation effect triggered by the tangential 
electric field boundary condition means that additional carbon particles, 
which could be introduced in the 'reaction zone' in the centre of the 
computational domain, would further enhance collisional discharges and 
lead to a denser tribo-electrically created plasma. Indeed, the charge 
localisation effect is the tribolectric plasma generation scenario as 
suggested by the Engineering Company Eco-Ardens
 experimental results (comp. with Fig.\ref{f2}). 
In these experiments, the additional carbon particles were brought in the 
tribo-plasma reaction zone by pyrolysis products.
In comparison with the tangential electric field boundary condition, the 
normal electric field has no significant effect on the localisation of 
particle charging. Hence, at least for the simulation run times attempted 
in this study, this other regime is not of interest from the point of 
view of tribo-electric plasma generation.
The rest of the section describes an analytical model for the stationary 
localised distribution of free charge accumulation in the case of the 
tangential electric field boundary condition.

Let us consider a two dimensional domain with periodic boundary conditions in the x- and y-direction. 
In comparison with the model considered in the previous section, in the 
present case the boundary problem is not homogeneous: the top and
the bottom boundaries in the y-direction correspond to conducting walls. 
On the walls, a periodic variation of the tangential electric field component 
is imposed, $E_x = E_x(x)$.
In accordance with the EPOCH solution (Fig. \ref{f10}), the electric field 
penetrates inside the domain and its effect decays away from the wall. 
To proceed with the analytical solution, let us model the effect of the non-homogeneous 
electric field on the ionized particle distribution by adding a diffusion term to the linear particle
collision model equation (3). At equilibrium
 $\partial n_1/\partial t = 0$ and the equation for the particle number per unit
volume becomes
\begin{equation}
\alpha n_e n_2  -\beta n_e n_1 + D \frac{\partial^2 n_1}{\partial y^2}=0,  
\end{equation}
where $D = const > 0$.
Let us discretise the solution domain into several non-overlapping bins in the y-direction where
the coordinates of each bin are
$0 \leq x \leq L_x, \,\,\, h \leq y \leq h+\delta h, \,\,\, 0 <h< L_y.$
By integrating Eq.(16) over each bin volume, one obtains
\begin{equation}
\alpha N_e \langle n_2 \rangle  -\beta N_e \langle n_1 \rangle + \langle D \frac{\partial^2 n_1}{\partial y^2} \rangle=0,  
\end{equation}
where the brackets mean averaging over the bin volume.
After a re-arrangement, using $n_1 + n_2 = n_0 $ and $\langle n_i \rangle = \langle n_1 \rangle / \langle n_0 \rangle$, 
$a=(\alpha+\beta)N_e$ and $b=\alpha/(\alpha+\beta)$, Eq(17) reduces to
\begin{equation}
\langle D \frac{d^2 n_i}{d y^2} \rangle= a \left(\langle n_i \rangle- b\right).  
\end{equation}
By introducing some effective average diffusion coefficient $\bar D$, the last equation can be
integrated to obtain
\begin{equation}
\langle n_i \rangle =b+A \exp \left( \sqrt{\frac{a}{\bar D}} y\right),  
\end{equation}
where $A = const$ is an amplitude parameter to be determined, e.g. from the boundary
condition.

At small distances from the bottom wall, $\sqrt{\frac{a}{\bar D}} y \ll 1$, Eq. (19) reduces to
integrated to obtain
\begin{equation}
\langle n_i \rangle =C + E y,  
\end{equation}
where $C = b=const$ and $E = A(1+\sqrt{\frac{a}{\bar D}}) = const$.

To close the model, the slope parameter $E$ in Eq.(20) can be related to the 
tangential electric field using the particle continuity and the
 electrostatic force equations as follows.
Let us consider the continuity equation for the number of ionized particles in a unit volume at
equilibrium:
\begin{equation}
\frac{\partial\left(u_x n_i\right)}{\partial x}+
\frac{\partial\left(u_y n_i\right)}{\partial y}=0. 
\end{equation}
Here $u_x$ and $u_y$ are effective x- and y- velocity components. The particle velocities are driven
by the non-homogeneous electric field.
By integrating Eq. (21) over the considered control volume close to the wall, to the first order, one
obtains
\begin{equation}
\frac{d \langle n_i\rangle}{d y} \approx -\langle n_i\rangle_{y=0}
{\langle \frac{\partial\left(u_x \right)}{\partial x}\rangle}/{U}. 
\end{equation}
Eq.(22) can be reduced to the form of Eq.(20),
where brackets correspond to the volume averaging and 
$E=-\langle n_i\rangle_{y=0}
{\langle \frac{\partial\left(u_x \right)}{\partial x}\rangle}/{U}$, 
which can be treated as constant to the first approximation.
Let us further approximate the particle velocity corresponding to 
their drift away from the wall by a constant value, $u_y = U > 0$ and 
take into account that the average number of particles does not depend on the x-coordinate.
To evaluate $
{\langle \frac{\partial\left(u_x \right)}{\partial x}\rangle}$ 
that appears as the slope, $E$, one can recall that the acceleration
exerted on a charged particle due to the electric field is given by
\begin{equation}
a_x (x, y) = qE_x (x, y)/m 
\end{equation}
where $q$ is the particle charge and $m$ is its mass, and using the 
standard kinematic relationships,
\begin{equation}
u_x=\int_0^t a_x (x, y)dt =\int_0^x \frac{a_x (x, y)}{u}dx \,\,\, {\rm and} \,\,\,
u \frac{\partial u }{\partial x}=a_x(x),  
\end{equation}
the integration over $[0, x]$, after some re-arrangement, leads to
\begin{equation}
u_x(x, y) =\sqrt{
\int_0^x {a_x (x, y)}dx +u_x^2(0,y)}.  
\end{equation}
Hence,
\begin{equation}
\langle \frac{d u_x}{dx}\rangle =\frac{1}{2}\langle \frac{a_x (x, y) -a_x (0, y)}
{\sqrt{\int_0^x {a_x (x, y)}dx +u^2(0,y)}}\rangle,  
\end{equation}
or
\begin{equation}
\langle \frac{d u_x}{dx} \rangle =\frac{q}{2m} 
\langle \frac{\Delta E_x}{\sqrt{\int_0^x {(q/m)E_x(x,y)}dx +u^2(0,y)}}\rangle,  
\end{equation}
where $\Delta E_x = E_x (x, y) -E_x (0, y)$.

The analytical model (20) can be compared with the output of the
EPOCH simulations which were provided in the form of the bin-averaged 
electron number normalised by the peak value $\langle n_e \rangle(y)/\langle n_e \rangle_{\rm max}$ 
as a function of the y-coordinate.
It can be first noted that
$\frac{\langle n_e \rangle}{\langle n_e \rangle_{\rm max}}=
1+\frac{\left(\langle n_e \rangle-\langle n_e \rangle_{\rm max}\right)}{\langle n_e \rangle_{\rm max}}$
and $\frac{\left(\langle n_e \rangle-\langle n_e \rangle_{\rm max}\right)}{\langle n_e \rangle_{\rm max}}\ll1$
in accordance.

\begin{figure}
\centering 
\includegraphics[width=0.5\textwidth]{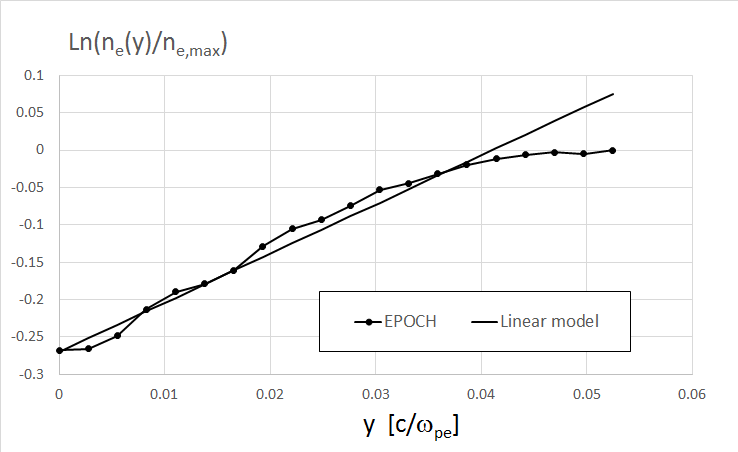}
\caption{ Inhomogeneous particle ionization problem: comparison of the EPOCH solution
(solid line with dots) with
the y-profile of the electron number density in 
 the linear model, according to Eq.(20). }
\label{f12}
\end{figure}

Hence, $\ln\left(\langle n_e \rangle(y)/\langle n_e \rangle_{\rm max} \right)
\propto \frac{\left(\langle n_e \rangle-\langle n_e \rangle_{\rm max}\right)}{\langle n_e \rangle_{\rm max}}
\propto \langle n_i \rangle$. The latter quantity is compared with Eq.(20) in 
Fig. \ref{f12}, where the two parameters of the linear model, $C$ and $E$ were 
selected from the best fit to the EPOCH data.
The good agreement between the fully kinetic plasma solution and the analytical 
model suggests that the assumptions used in the model are reasonable for the 
triboelectric plasma generation regime of interest. 

\section{Conclusions}

In this work we present PIC simulations of free 
charge creation by collisional ionization of C12 and C60 particles in 
plasma for the parameters of relevance to plasma gasification. 
For plasma simulations a fully collisional EPOCH model is used and the obtained 
solutions are reasonably non-sensitive to the numerical parameters such as 
the grid resolution, the domain size and the PPC number. 
There are two regimes considered: with and without excitation of the 
non-uniform electric field on the boundary.
Our main findings are as follows:

(i) In uniform plasmas with smooth walls there appear to be two 
optimal values of $C60:C12$ fraction for free electron production by 
collisional ionization (i.e. a most efficient discharge condition creation): 
one is 
$10:90$ and the other is $80:20$. The first value is in agreement with the 
experimental results of LCC Engineering who performed gasification tests 
with relatively low fullerene concentrations.

(ii) In plasmas with rough walls, modelled by comb-like electric field 
distribution at the boundary, the case of tangential electric field creates a 
significant charge localization in C12+ and C60+ species. This leads to the most favorable 
discharge condition creation for tribo-electrically generated plasma.

(iii) Linear analytical models are presented for modelling the particle 
collision process. Predictions of the models are in an encouraging 
agreement with the numerical simulation results.





\begin{acknowledgments}
This research utilized Queen Mary University of London's (QMUL) 
MidPlus computational facilities,       
supported by QMUL Research-IT.
\end{acknowledgments}


\begin{figure}
\centering
\includegraphics[width=3.5cm]{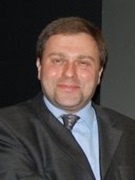}
\caption{ David Tsiklauri was born in Tbilisi, Georgia, in 1972. 
 He received the M.Sci. degree in theoretical physics from Tbilisi State University, Tbilisi, in 1994, 
 and the Ph.D. degree in physics from the University of Cape Town, Cape Town, South Africa, in 1996.
He was a Postdoctoral Fellow with the University of Cape Town, Tbilisi State University, 
Iowa State University, Ames, IA, USA, and University of Warwick, Coventry, U.K. 
From 2003 to 2009, he was a Lecturer and a Reader with the University of Salford, Greater Manchester, U.K. 
In 2009, he joined the Queen Mary University of London, where he is currently a Senior Lecturer. 
His current research interests include novel particle accelerator concepts, plasma wake-field acceleration, 
enhanced dissipation of MHD waves in inhomogeneous plasmas, collisionless magnetic reconnection, 
particle acceleration by dispersive Alfven waves, and radio emission generation mechanisms 
by accelerated electrons. }
\end{figure}

\begin{figure}
\centering
\includegraphics[width=3.5cm]{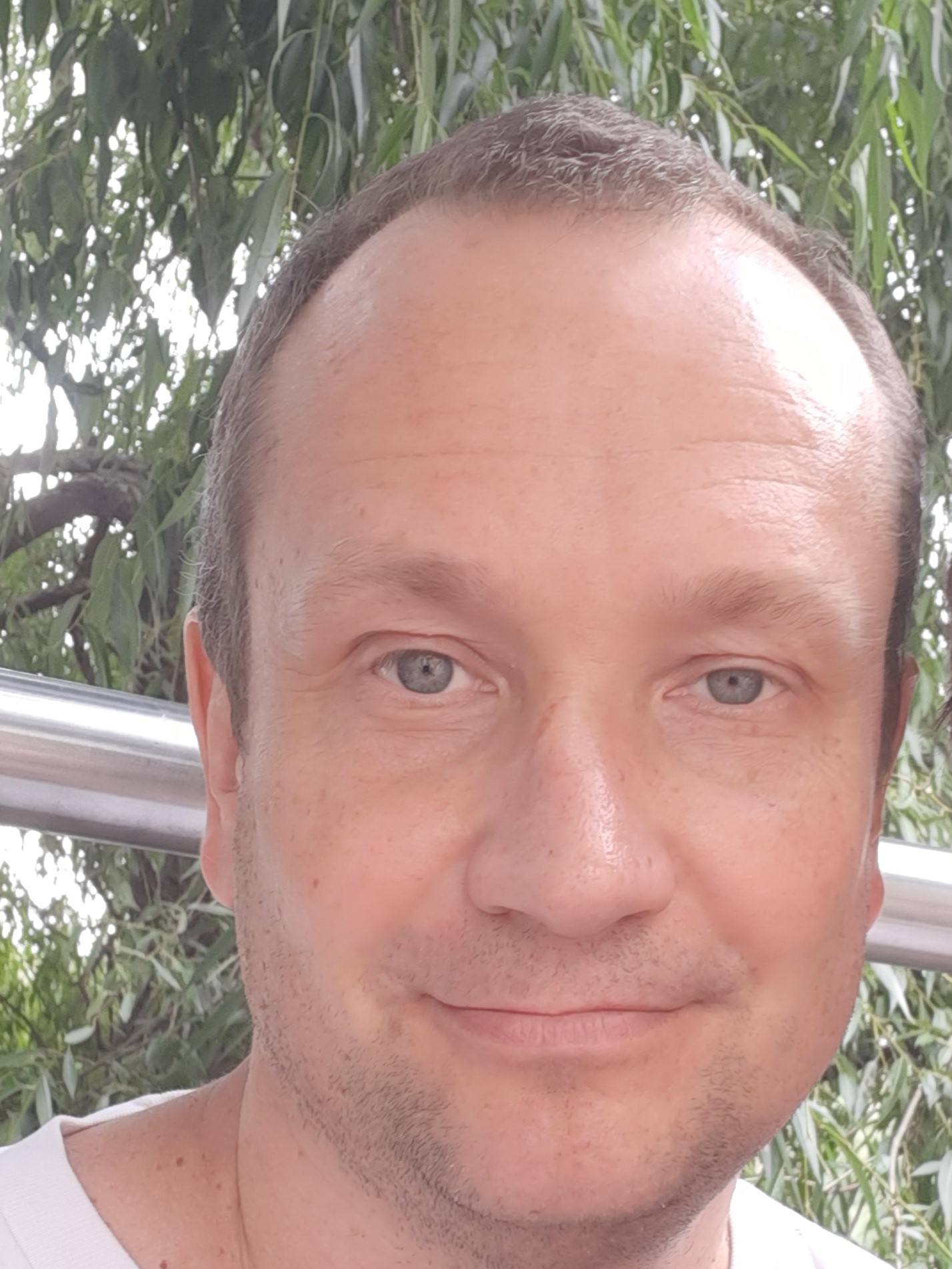}
\caption{Sergey Karabasov is Reader in Computational Modelling in Queen Mary University of 
London, School of Engineering and Materials Science. Sergey obtained MSc in 
applied mathematics and physics from Moscow Institute of Physics and Technology 
in 1995 and PhD in Mathematical Modelling from Moscow Lomonosov State University in 1999.
Prior to joining Queen Mary as a Senior Lecturer in 2012, he was a Royal Society 
University Research Fellow in Cambridge University Engineering Department developing 
hybrid methods for computational aeroacoustics. In 2010, Sergey was awarded the Full 
Doctorate of Science (Habilitation Degree) from Keldysh Institute of Applied Mathematics, 
Moscow for his work on hybrid and direct models in computational aeroacoustics. 
Sergey's work typically combines high-resolution computational methods (large-eddy 
simulations, numerical schemes, high-performance computing) with model decomposition 
approaches in aero- and hydro-dynamics (acoustic analogy, integral surface methods, 
vortex sound theory, ocean modelling) and multiscale modelling 
(bridging continuum mechanics with atomistic simulations).}
\end{figure}

\begin{figure}
\centering
\includegraphics[width=3.5cm]{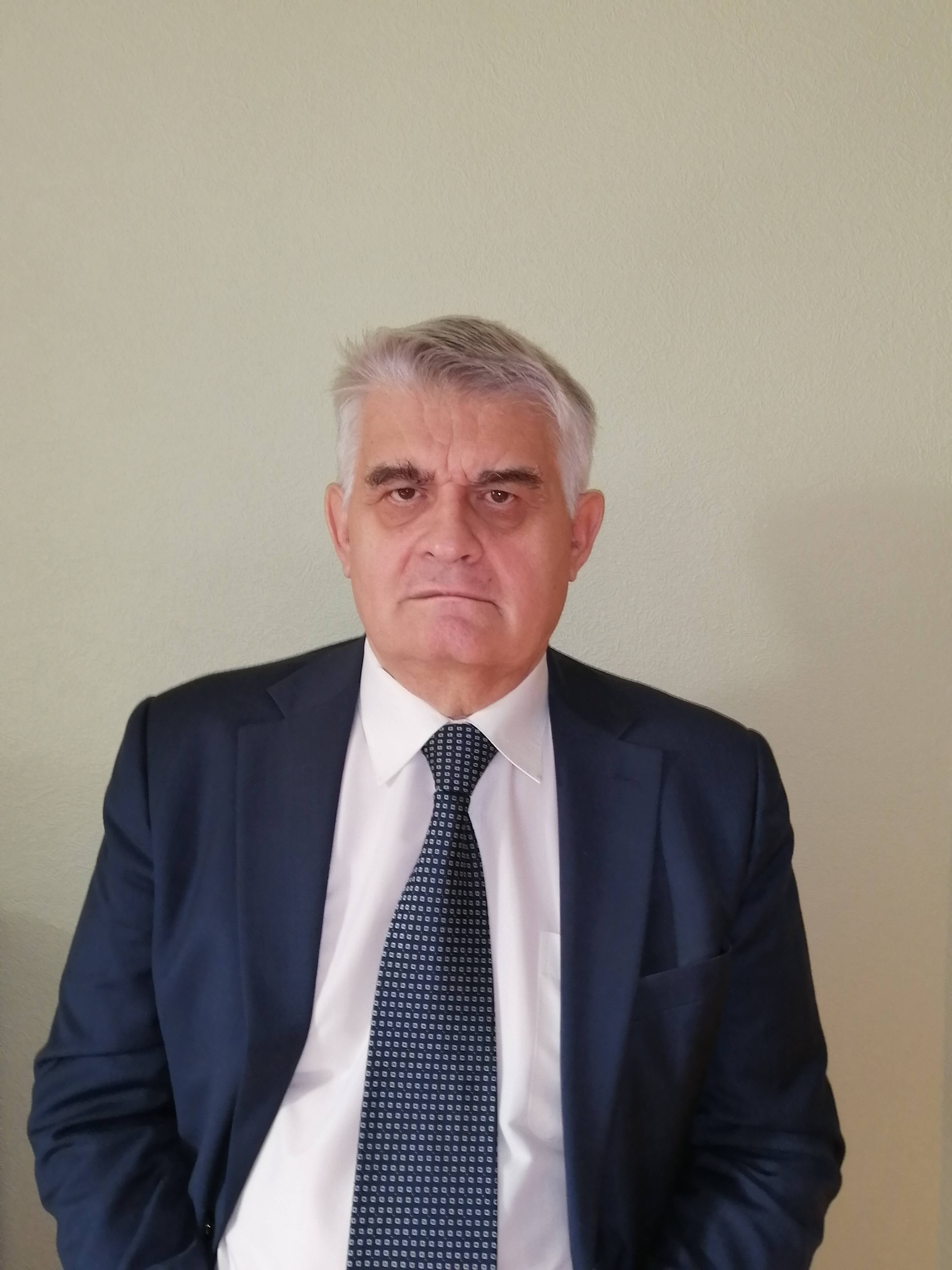}
\caption{Vladimir Grigorievich Prodaevich graduated from the 
Mari Polytechnic Institute in 1980. He worked in forestry in the 
Nizhny Novgorod region until 1992. In 1992 he was appointed Director 
General of the Russian Forest Association. He developed a technology for the 
utilization of waste from the Balakhna PPM with the production of highly 
effective biologically active substances. From 1994 to 1999, he was the 
Chairman of the Board of Directors of the Dubitel plant in Vyshny Volochyok. 
During these years he developed a technology for obtaining new types of products 
from spruce and willow bark. He developed a preparation for clarification of 
wine materials and juices as well as a technology for the production of 
environmentally friendly wood-based panels for housing construction. 
He is also an author of a preparation for increasing of the strength of the 
hermetic layer of tubeless car tires. When leading the Central Research Forest 
Chemical Institute (TsNILHI) he developed a technology for producing bitulin from 
birch bark. He also developed a technology of complex processing of spruce 
bark to produce lipids, tannides, pectins in a single technological process. 
The developments include a continuous retort for the production of charcoal with the 
utilization of exhaust gases using a plasma gasifier. Since 2009 he has been 
working on the improvement of the design of a plasma gasifier for various purposes. 
His recent work on the design of tribo-electric plasma gasifiers has been 
with Engineering Company Eco-Ardens.}
\end{figure}

\end{document}